%% file: ms.tex
\documentclass[iop, numberedappendix]{emulateapj} 

\usepackage{graphicx}
\usepackage{pslatex}
\usepackage{amsmath, amsthm, amssymb}
\usepackage{natbib}
\usepackage{epsfig}
\usepackage{subfigure}
\usepackage{enumerate}
\usepackage{upgreek}
\usepackage{accents}
\usepackage{ulem}
\usepackage{color}

\shorttitle{Global 21-cm Parameter Estimation}
\shortauthors{Mirocha et al.}

\begin{document}

%%% Custom definitions
\input{macros.tex}

%%% 
%% Title and Affiliations
%%%
\title{Interpreting the Global 21-cm Signal from High Redshifts. \\ 
II. Parameter Estimation for Models of Galaxy Formation}
\author{Jordan Mirocha\altaffilmark{1,$\dagger$,$\star$}, Geraint J.A. Harker\altaffilmark{2,$\ddagger$}, Jack O. Burns\altaffilmark{1,3}}

\altaffiltext{1}{Center for Astrophysics and Space Astronomy and Department of Astrophysical and Planetary Science, University of Colorado, Campus Box 389, Boulder, CO 80309}
\altaffiltext{2}{Department of Physics and Astronomy, University College London, London WC1E 6BT, UK}
\altaffiltext{3}{NASA Ames Research Center, Moffet Field, CA 94035, USA}
\altaffiltext{$^{\dagger}$}{NASA Earth and Space Science Graduate Fellow}
\altaffiltext{$^{\star}$}{Now at UCLA; mirocha@astro.ucla.edu}
\altaffiltext{$^{\ddagger}$}{Marie Curie Fellow}

%%%
%% Abstract
%%%
\begin{abstract}
Following our previous work, which related generic features in the
sky-averaged (global) 21-cm signal to properties of the intergalactic medium,
we now investigate the prospects for constraining a simple galaxy formation
model with current and near-future experiments. Markov-Chain Monte Carlo fits
to our synthetic dataset, which includes a realistic galactic foreground, a
plausible model for the signal, and noise consistent with 100 hours of
integration by an ideal instrument, suggest that a simple four-parameter model
that links the production rate of Lyman-$\alpha$, Lyman-continuum, and X-ray
photons to the growth rate of dark matter halos can be well-constrained (to
$\sim 0.1$ dex in each dimension) so long as all three spectral features
expected to occur between $40 \lesssim \nu / \mathrm{MHz} \lesssim 120$ are
detected. Several important conclusions follow naturally from this basic
numerical result, namely that measurements of the global 21-cm signal can in
principle (i) identify the characteristic halo mass threshold for star
formation at all redshifts $z \gtrsim 15$, (ii) extend $z \lesssim 4$ upper
limits on the normalization of the X-ray luminosity star-formation rate
($L_X$-SFR) relation out to $z \sim 20$, and (iii) provide joint constraints
on stellar spectra and the escape fraction of ionizing radiation at $z \sim
12$. Though our approach is general, the importance of a broad-band
measurement renders our findings most relevant to the proposed \textit{Dark
Ages Radio Explorer}, which will have a clean view of the global 21-cm signal
from $\sim 40-120$ MHz from its vantage point above the radio-quiet,
ionosphere-free lunar far-side.
\end{abstract}
\subjectheadings{early universe --- diffuse radiation -- epoch of reionization}

%%%
%% Introduction
%%%
\section{Introduction} \label{sec:Introduction}
The high redshift ($z \gtrsim 6$) Universe has become a frontier in recent
years, as it was the time in which stars, galaxies, and black holes first
formed, bringing an end to the cosmic ``dark ages'' and initiating the
``cosmic dawn.'' Preliminary searches for these objects have commenced,
primarily with the \textit{Hubble Space Telescope}, and have begun to find
very bright galaxies at $z \sim 10$ \citep{Zheng2012,Ellis2013,Oesch2013}.
However, it is the individually faint but overwhelmingly numerous galaxies
that likely usher in the Epoch of Reionization (EoR), and may account for a
substantial fraction of the ionizing photons required to bring the EoR to a
close by $z \sim 6$ \citep[e.g.,][]{Trenti2010,Wise2014}. If the galaxy
luminosity function flattens considerably at the low-luminosity end
\citep{OShea2015} or the formation efficiency of massive Population III stars
is low, even the \textit{James Webb Space Telescope} may struggle to find
faint galaxies beyond $z \sim 10$ \citep{Zackrisson2012}. Measurements of the cosmic microwave background (CMB), in conjunction with independent constraints on the ionization and star-formation histories, support a relatively short EoR, and thus a relatively modest galaxy population at $z \gg 10$ \citep{Planck2015,Robertson2015,Bouwens2015}.

Observations at low radio frequencies -- corresponding to highly redshifted
21-cm ``spin-flip'' radiation from neutral hydrogen atoms -- are a promising
complement to CMB measurements and optical and near-infrared imaging campaigns to constrain the
high redshift galaxy population
\citep{Madau1997,Shaver1999,FurlanettoOhBriggs2006}. Numerous efforts to
detect the 21-cm background are already underway, including both its spatial
fluctuations and monopole spectral signature. For instance, observations with
the \textit{Precision Array for Probing the Epoch of Reionization} indicate an
X-ray heated IGM at $z \sim 7.7-8.4$ \citep{Parsons2014,Ali2015,Pober2015},
while the \textit{Experiment to Detect the Global EoR Signal} (\edges) is so
far the only 21-cm experiment to set lower limits on the duration of the
reionization epoch \citep[$\Delta z > 0.06$;][]{Bowman2010}.

Targeting the sky-averaged (``global'') 21-cm spectrum may enable more rapid
progress at the highest redshifts ($z \gtrsim 10$), as it can in principle be
detected with a single well-calibrated dipole receiver. Several ground-based
experiments are underway \citep[e.g.; EDGES, SCI-HI, LEDA,
BIGHORNS;][]{Bowman2010,Voytek2014,Greenhill2012,Sokolowski2015}, though
space-based observatories will be required to probe the cosmic dawn at $z
\gtrsim 30$ \citep[e.g., the \textit{Dark Ages Radio Explorer},
(DARE);][]{Burns2012}, as the Earth's ionosphere reflects and refracts radio
signals at low frequencies \citep{Vedantham2013,Datta2015}. Interferometers
are in principle capable of measuring the global 21-cm signal, so long as they
are compact \citep{Presley2015,Singh2015} or can overcome challenges in lunar
occultation techniques \citep{Vedantham2014}.

No matter the observing technique, all global experiments must cope with the
Galactic foreground, which is $\sim 10^4-10^6$ times brighter than the
cosmological signal in temperature. Though strong and spatially variable, the Galactic
foreground is spectrally smooth \citep{deOliveiraCosta2008}, in contrast to
the expected high redshift signal which is spatially invariant but spectrally complex. It is this spectral structure that should enable one to
distinguish foreground from signal
\citep{Shaver1999,Gnedin2004,Furlanetto2006,Pritchard2010a}, especially if one
observes multiple (semi-) independent sky regions
\citep{Harker2012,Liu2013,Switzer2014}. Though current constraints do not rule
out unresolved spectral structure at the level of the high redshift signal,
theoretical arguments favor a smooth foreground at the relevant frequencies
\citep{Petrovic2011,Bernardi2015}. Extragalactic point sources are another
foreground, but for experiments with broad beams, their combined contribution
averages into another diffuse spectrally smooth foreground \citep{Shaver1999}.

Given that the global 21-cm signal is an indirect probe of high-$z$ galaxies,
some modeling is required to convert observational quantities to constraints
on the properties of the Universe's first galaxies. Though numerous studies
have performed forward modeling to predict the strength of the global 21-cm
signal \citep{Furlanetto2006,Pritchard2010a}, few have attempted to infer
physical parameters of interest from synthetic datasets. Such forecasting
exercises are incredibly useful tools for designing instruments and planning
observing strategies, as they illuminate the mapping between constraints on
observable quantities and model parameters of interest. Both Fisher matrix and
Markov Chain Monte Carlo (MCMC) approaches have been employed by the power
spectrum community \citep[e.g.,][]{Pober2014,Greig2015}, the latter providing
a powerful generalization that does not require the assumption of Gaussian
errors or perfect recovery of the maximum likelihood point. 

Most work to date has instead focused on forecasting constraints on
\textit{phenomenological} parameters of interest, e.g., the timing and
duration of reionization \citep{Liu2013}, the depth and width of the deep
minimum expected near $\sim 70$ MHz prior to reionization
\citep{Bernardi2015}, or all three spectral features predicted to occur
between $40 \lesssim \nu \ / \mathrm{MHz} \lesssim 120$
\citep{Pritchard2010a,Harker2012}. These spectral ``turning points'' in the
global 21-cm spectrum provide a natural basis for parameter forecasting, as
they persist over large ranges of parameter space \citep{Pritchard2010a}, can
be extracted from the foreground with realistic instruments and integration
times \citep[at least under the assumption of a negligible
ionosphere;][]{Harker2012,Presley2015,Bernardi2015}, and can be extracted from
the foreground even when their positions are not used as the parameters of a
signal model \citep{Harker2015b}. They can also be interpreted fairly robustly
in terms of the physical properties of the IGM, at least in simple two-phase
models \citep[][hereafter Paper I]{Mirocha2013}. Given the viability of the
turning points as ``products'' of global 21-cm signal extraction pipelines, we
will use them as a launching point in this paper from which to explore the
prospects for constraining \textit{astrophysical} parameters of interest with
observations of the global 21-cm signal. Importantly, we will consider all
three turning points simultaneously, rendering our findings particularly
applicable to \dare, whose band extends from $40 \leq \nu / \mathrm{MHz} \leq
120$ in order to maximize the likelihood of detecting all three features at
once.

This paper is organized as follows. In Section 2 we outline our methods for modeling the global 21-cm signal and parameter estimation. Section 3 contains our main results, with a discussion to follow in Section 4. In Section 5, we summarize our results. We use the most up-to-date cosmological parameters from \textit{Planck} throughout \citep[last column in Table 4 of][]{Planck2015}.

%%%
%% METHODS
%%%
\section{NUMERICAL METHODS}
In order to forecast constraints on the properties of the first galaxies, we
will need (1) a model for the global 21-cm signal, (2) estimates for the
precision with which this signal can be extracted from the foregrounds, and
(3) an algorithm capable of efficiently exploring a multi-dimensional
parameter space to find the best-fit model parameters and their uncertainties.
The next three sub-sections are devoted to describing these three pieces of
our pipeline in turn.

%%
% PHYSICAL MODEL
%%
\subsection{Physical Model for the Global 21-cm Signal} \label{sec:physical_model}
Our approach to modeling the global 21-cm signal is similar to that presented
in several other published works \citep[e.g.,][]{Barkana2005,Furlanetto2006,
Pritchard2010a,Mirocha2014}, so we will only discuss it here briefly. The
primary assumption of our model is that the radiation backgrounds probed by
the turning points are generated by stars and their byproducts, which form at
a rate proportional to the rate of baryonic collapse into dark matter haloes.
That is, we model the star-formation rate density (SFRD) as
\begin{equation}
    \rhostardot(z) = \fstar \rhobbar \frac{d \fcoll}{dt} ,
\end{equation}
where $\fcoll=\fcoll(\Tmin)$ is the fraction of matter in collapsed halos with
virial temperatures greater than $\Tmin$, $\rhobbar$ is the mean baryon
density today, and $\fstar$ is the star formation efficiency. We use a fixed
$\Tmin$ rather than a fixed $M_{\min}$ because it provides physical insight
into the processes governing star-formation, as one can easily identify the
atomic and molecular cooling thresholds of $\sim 10^4$ and $\sim 500$ K. Note
that a fixed value of $\Tmin$ results in a time-dependent mass threshold,
$M_{\min}$.

In order to generate a model realization of the global 21-cm signal, we must
convert star-formation to photon production. Given that the three spectral
turning points probe the history of ionization, heating, and $\Lya$ emission,
we will split the production of radiation into three separate bands: (1) from
the $\Lya$ resonance to the Lyman-limit, $10.2 \leq h\nu /
\mathrm{eV} \leq 13.6$, which we refer to as the Lyman-Werner
(LW) band despite its inclusion of photons below $11.2$ eV, (2)
hydrogen-ionizing photons, with energies $13.6 \leq h\nu /
\mathrm{eV} \leq 24.4$, and (3) X-rays, with energies exceeding 0.1 keV. Each
radiation background is linked to the SFRD through a scaling parameter $\xi$, which converts a rate of star formation to a rate of photon (or energy) production). The $\Lya$ background intensity is then given by
\begin{align}
    \Jalpha(z) = \frac{c}{4\pi} \frac{(1+z)^2}{H(z)} \frac{\xi_{\mathrm{LW}}}{\Delta \nu_{\alpha}} \rhobbar \frac{d \fcoll}{dt}  , 
\end{align}
where $\Delta \nu_{\alpha} = \nu_{\mathrm{LL}} - \nu_{\alpha}$, and the ionization rate by
\begin{equation}
    \Gamma_{\HI} = \xi_{\ioniz} \rhobbar \frac{d \fcoll}{dt}
\end{equation}
The rate of X-ray heating is defined instead in terms of an energy per unit star-formation, i.e.,
\begin{equation}
    \eX = f_{X,h} c_X \xi_X \rhobbar \frac{d \fcoll}{dt}
\end{equation}
where $c_X$ is the normalization of the $L_X$-SFR relation, which we take to be $c_X = 3.4 \times 10^{40} \ \cXunits$ following \citet{Furlanetto2006}\footnote{\citet{Furlanetto2006} computed this value by extrapolating the 2-10 keV $L_X$-SFR relation of \citet{Grimm2003} above 0.2 keV, assuming an unabsorbed $\alpha=1.5$ power-law spectrum. Our reference value of $\xi_X = \fstar f_X = 0.02$ (which is dimensionless, unlike $\xiLW$ and $\xiUV$) is chosen to match recent analyses in the 0.5-8 keV band, which find $c_X = 2.6 \times 10^{39} \ \cXunits$ \citep{Mineo2012a}}.
Note that we have absorbed $\fstar$ into the $\xi$ parameters, i.e.
\begin{align}
    \xi_{\mathrm{LW}} & = \Nlw \fstar \label{eq:xiLW} \\
    \xi_{\mathrm{ion}} & = \Nion \fstar \fesc  \label{eq:xiUV} \\
    \xi_X & = \fstar f_X,
\end{align}
where $\Nlw$ and $\Nion$ are the number of LW and ionizing photons emitted per stellar baryon, $\fesc$ is the escape fraction of ionizing radiation, and $f_X$ scales the $L_X$-SFR relation.

Given $\eX$, $\Gamma_{\HI}$, and the $\Lya$ background intensity, $\Jalpha$, we can evolve the ionization and thermal state of intergalactic gas in time, and compute the
sky-averaged 21-cm signal, i.e., the differential brightness temperature of HI relative to the CMB, via \citep[e.g.,][]{Furlanetto2006}\footnote{Though this expression does not explicitly depend on the density of hydrogen gas (it assumes gas is at the cosmic mean density, $\delta = 0$), the global 21-cm signal may still be sensitive to the density provided that fluctuations in the ionization and/or spin temperature of HI gas are correlated with its density.}
\begin{equation}
    \dTb \simeq 27 (1 - \xibar) \left(\frac{\Obnow h^2}{0.023} \right) \left(\frac{0.15}{\Omnow h^2} \frac{1 + z}{10} \right)^{1/2} \left(1 - \frac{\Tcmb}{T_{\mathrm{S}}} \right) , \label{eq:dTb}
\end{equation}
where $\Tcmb$ is the CMB temperature, $\xibar$ is the volume-averaged ionization fraction, 
\begin{equation}
    \xibar = \QHII + (1 - \QHII) x_e
\end{equation} with $\QHII$ representing the volume-filling factor of HII
regions and $x_e$ the ionized fraction in the bulk IGM. $T_S$ is the
excitation or ``spin'' temperature of neutral hydrogen, which quantifies the
number of hydrogen atoms in the hyperfine triplet and singlet states,
\begin{equation}
    T_S^{-1} \approx \frac{T_{\gamma}^{-1} + x_c T_K^{-1} + x_{\alpha} T_{\alpha}^{-1}}{1 + x_c + x_{\alpha}}
\end{equation}
where $T_K$ is the gas kinetic temperature, $T_{\alpha} \simeq T_K$
\citep{Field1958}, and $h$ and the $\Omega$'s take on their usual cosmological
meaning. We compute the collisional coupling coefficient, $x_c$, by
interpolating between the tabulated values in \citet{Zygelman2005} with a
cubic spline, and take $x_{\alpha} = 1.81 \times 10^{11} \Jalpha
/ (1 + z)$. We perform these calculations using the Accelerated Reionization
Era Simulations (\textsc{ares})
code\footnote{\url{https://bitbucket.org/mirochaj/ares; v0.1}},
which is the union of a 1-D radiative transfer code developed in
\citet{Mirocha2012} and uniform radiation background code described in
\citet{Mirocha2014}. See \S2 of \citet{Mirocha2014} for a more detailed
description of the global 21-cm signal modeling procedure.

\begin{figure*}[htbp]
\begin{center}
\includegraphics[width=0.8\textwidth]{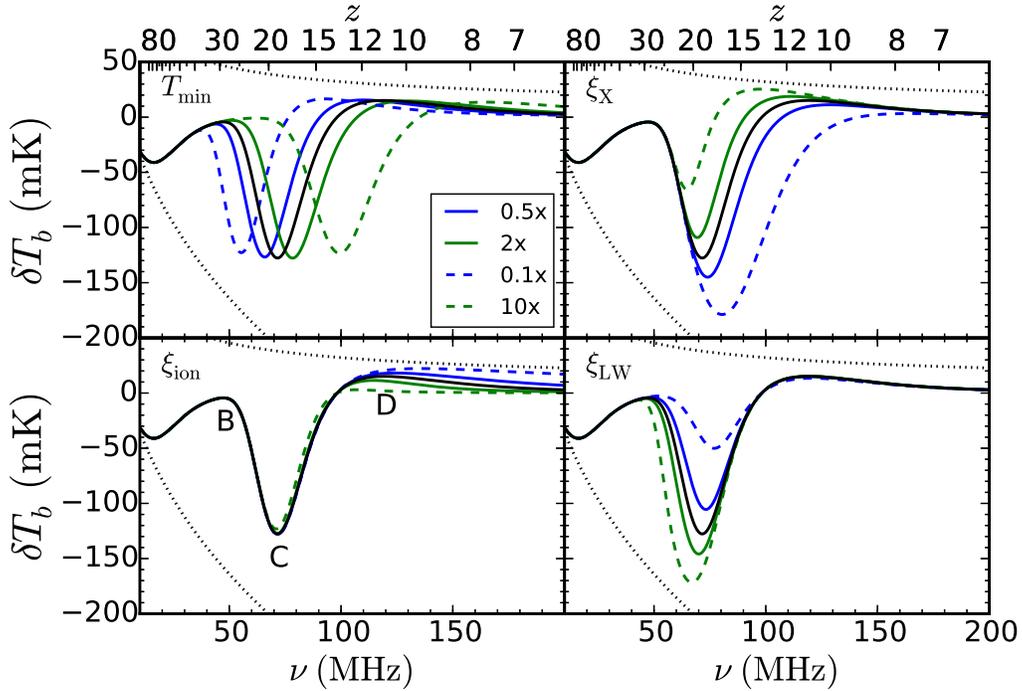}
\caption{Illustration of the basic dependencies of the global 21-cm signal. The black line is the same in each panel, representing our reference model (see Tables \ref{tab:reference_model} and \ref{tab:priors}), while all solid green (blue) lines correspond to a factor of 2 increase (decrease) in the parameter noted in the upper left corner, and dashed lines are factor of 10 changes above and below the reference value. The right half of the figure is qualitatively similar to Figure 2 of \citet{Pritchard2010a}, though our reference values for the $\xi$ parameters are different, as are our cosmological parameters, leading to quantitative differences. The dotted lines show the maximum allowed amplitude of the signal (i.e., the amplitude of the signal when $\xibar = 0$ and $T_S >> T_{\gamma}$), and the minimum allowed amplitude of the signal (set by assuming $T_S = T_K = T_{\mathrm{ad}}(z)$, where $T_{\mathrm{ad}}$ is the gas temperature in an adiabatically-cooling Universe). Because we refer to the spectral features as turning points B, C, and D throughout the paper, we annotate them in the lower left panel for reference.}
\label{fig:param_study}
\end{center}
\end{figure*}

\begin{deluxetable*}{lllll}
\tabletypesize{\scriptsize}
\tablecaption{Reference Model Properties and Simulated Constraints}
\tablecolumns{5}
\tablehead{Quantity & Reference Value & EM1 & EM2 & Description} 
\startdata
$\nu_\mathrm{B} \ [\mathrm{MHz}]$ & 47.4 & $46.99 \pm 0.74$ & $47.08 \pm 0.60$ & Onset of $\Lya$ coupling \\
$\nu_\mathrm{C} \ [\mathrm{MHz}]$ & 71.0 & $70.95 \pm 0.20$ & $70.96 \pm 0.15$ & Onset of heating \\
$\nu_\mathrm{D} \ [\mathrm{MHz}]$ & 111.4 & $110.9 \pm 5.0$ & $109.2 \pm 3.5$ & Beginning of reionization \\
$\dTb(\nu_\mathrm{B}) \ [\mathrm{mK}]$ & -4.4 & n/a & n/a & Depth when $\Lya$ coupling begins \\
$\dTb(\nu_\mathrm{C}) \ [\mathrm{mK}]$ & -124.8  & $-122.6 \pm 5.0$ & $-121.7 \pm 3.7$ & Depth of absorption trough \\
$\dTb(\nu_\mathrm{D}) \ [\mathrm{mK}]$ & 19.2   & $17.20 \pm 4.5$ & $19.88 \pm 1.7$ & Height of emission feature \\
\hline
$\zrei$ & 9.25 & n/a & n/a & Midpoint of reionization \\
$\tau_e$ & 0.066 & n/a & n/a & CMB optical depth
\enddata
\tablecomments{Observational properties of our reference model (solid black lines in Figure \ref{fig:param_study}), and the best-fit and uncertainties for each extraction model (EM) we consider. Subscripts indicate different turning points, i.e., the cosmic dawn feature when the Wouthuysen-Field effect first  drives $T_S$ to $T_K$ (B), the absorption trough, which indicates the onset of heating (C), and the beginning of reionization (D). EM1 and EM2 differ in the number of independent sky regions assumed (1 vs. 2), and in the complexity of the foreground model (3rd vs. 4th order polynomial), which leads primarily to a more robust detection of turning point D for EM2. All errors are $1-\sigma$, and correspond directly to the diagonal elements of the turning point covariance matrix.}
\label{tab:reference_model}
\end{deluxetable*}

\begin{figure}[htbp]
\begin{center}
\includegraphics[width=0.48\textwidth]{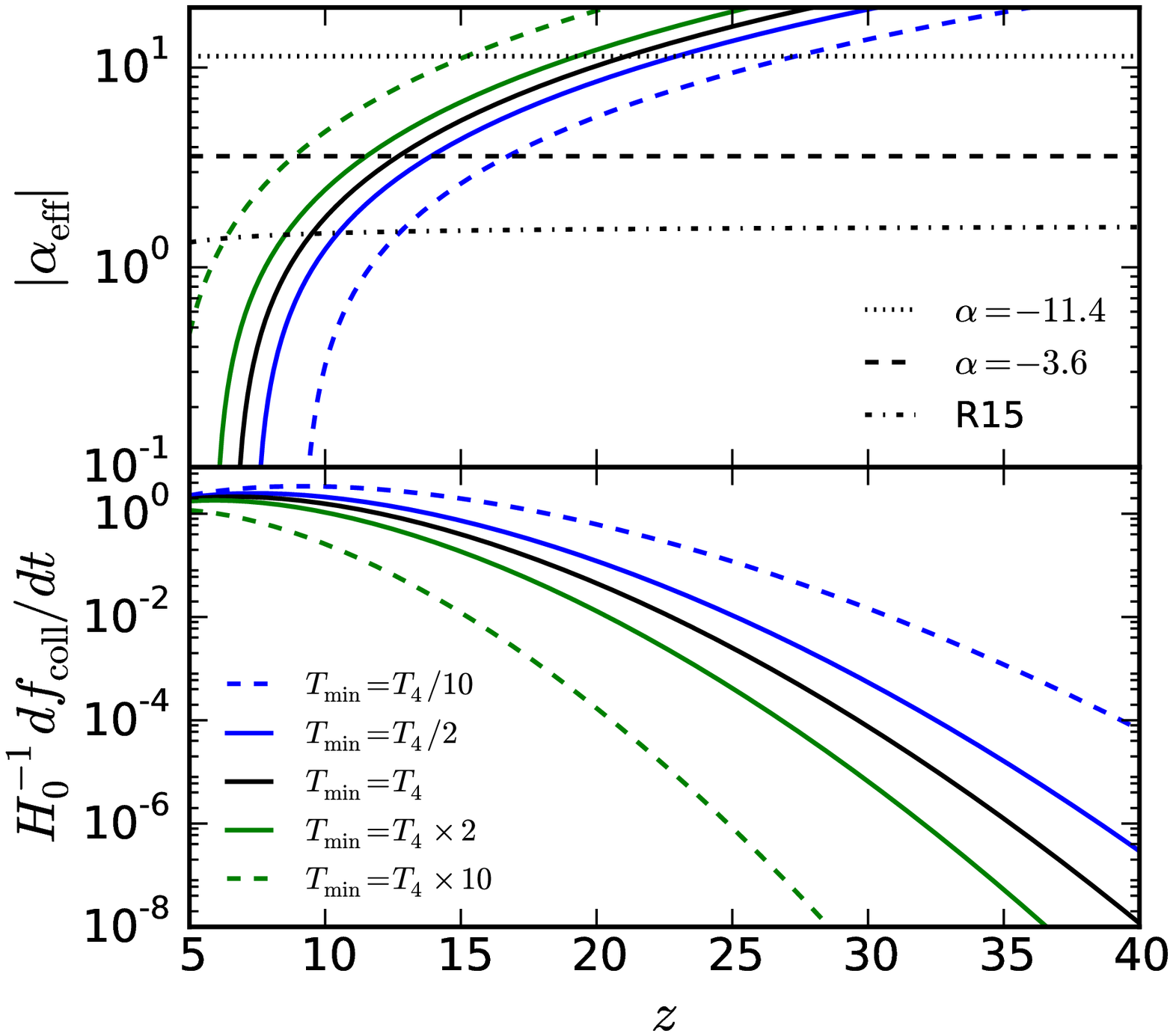}
\caption{\textit{Bottom:} Rate of collapse onto halos above a given virial temperature threshold, $\Tmin$, scaled to the Hubble time. \textit{Top:} Effective power-law index (Eq. \ref{eq:alpha_eff}) as a function of redshift for each $d\fcoll/dt$ model. Empirical power-laws from \citet{Oesch2012} are overlaid for comparison, as well as the best-fit 4-parameter SFRD model used in \citet{Robertson2015}.}
\label{fig:fcoll_2panel}
\end{center}
\end{figure}

Figure \ref{fig:param_study} shows our reference model (properties of which
are listed in Table \ref{tab:reference_model}), and the modulations in its
structure that occur when varying $\Tmin$, $\xi_X$, $\xi_{\ioniz}$, and
$\xi_{\mathrm{LW}}$. It is immediately clear that $\Tmin$ affects the
locations of all three turning points, whereas each $\xi$ parameter affects at
most two. We should therefore expect that in principle, an observation
containing all three features will have the best chance to constrain $\Tmin$,
though this will be complicated at the lowest redshifts where $d\fcoll/dt$
becomes a weaker function of $\Tmin$ (see bottom panel of Figure
\ref{fig:fcoll_2panel}).

Figure \ref{fig:param_study} also shows that $\xi_{\ioniz}$ will be difficult
to constrain using global signal data at these frequencies, as even factor of
10 changes lead only to small changes in the signal (at $\nu \gtrsim 100$
MHz), whereas factor of 10 changes in $\xi_X$ and $\xi_{\mathrm{LW}}$ are
$\sim 50$ mK effects. There are also clear degeneracies between $\Tmin$ and
the $\xi$ parameters. Exploring those degeneracies and determining the
prospects for constraining each parameter independently are our primary goals
in this work. The results will in large part depend on how accurately the
signal can be recovered from the foregrounds, which we discuss in the next
subsection.

Before moving on to signal recovery, it is worth reiterating that our approach
cannot be used to constrain the normalization of the SFRD, since we have
absorbed the star-formation efficiency into the $\xi$ parameters. However, we
can constrain the rate-of-change in the SFRD, as it is uniquely determined by
$\Tmin$. It is illustrative to quantify this using an effective power-law
index
\begin{equation} 
    \alpha_{\mathrm{eff}}(z) \equiv \frac{d\log \rhostardot(z)}{d\log(1+z)} , \label{eq:alpha_eff}
\end{equation} 
which enables a straightforward comparison with empirical models, which are
often power-laws, i.e., $\rhostardot(z) \propto (1 + z)^{\alpha}$, in which
case $\alpha = \alpha_{\mathrm{eff}} = \mathrm{constant}$. The
$\alpha_{\mathrm{eff}}(z)$ values of our $\fcoll$ model are independent of
$\fstar$ so long as $d\fstar/dt = 0$, and generally fall within the (broad)
range of values permitted by observations of high-$z$ galaxies
\citep{Oesch2013,Robertson2015}, as shown in the top panel of Figure
\ref{fig:fcoll_2panel}.

%%
% "OBSERVATION"
%%
\subsection{Signal Extraction} \label{sec:mockobs}
In order to fit a physical model to the turning points of the global 21-cm
signal, we require best-fit values for the turning point positions and
estimates for uncertainties. To do this, we build on the work of
\citet{Harker2012} and \citet{Harker2015b}, who introduced a MCMC technique for fitting global 21-cm signal data. The basic
approach is to simultaneously fit a model for the galactic foreground, the
global 21-cm signal, and in general, parameters of the instrument (e.g., its
response as a function of frequency), assuming some amount of integration
time, $\tint$, and the number of independent sky regions observed, $\Nsky$.
The foreground is modeled as a polynomial in $\log \nu - \log T$ space, while
the astrophysical signal is modeled as either a spline \citep{Harker2012} or
series of $\tanh$ functions,
\begin{equation}
  A(z)=\frac{A_{\mathrm{ref}}}{2}\{1+\tanh[(z_0-z)/\Delta z]\}\ ,
  \label{eq:tanhdef}
\end{equation}
that represent $\Jalpha(z)$, $T_K(z)$, and $\xibar(z)$ \citep{Harker2015b}. The free parameters of the tanh model are the ``step height,'' $A_{\mathrm{ref}}$, pivot redshift, $z_0$, and a duration, $\Delta z$. 

The tanh approach to modeling the global 21-cm signal was chosen for
numerous reasons. First and foremost, it was chosen as a computationally
efficient substitute for more expensive, but physically-motivated models like
those investigated in this paper. Some alternative intermediaries include the
`turning points' parameterization \citep{Pritchard2010a,Harker2012} or models
that treat the absorption feature as a Gaussian \citep{Bernardi2015}. Both are
comparably cheap computationally, but cannot capture the detailed shape of
physical models. Perhaps most importantly, the spline and Gaussian models are
purely phenomenological, making them difficult to interpret in terms of IGM or
galaxy properties and thus incapable of incorporating independent prior
information on e.g., the ionization or thermal history. The $\tanh$ approach,
on the other hand, can mimic the shape of typical global 21-cm signal models
extremely well, and can be immediately related to physical properties of the
IGM. 

\citet{Harker2015b} presented a suite of calculations spanning the
2-D parameter space defined by $\Nsky = \{1,2,4,8\}$ and $\tint = \{100,
1000\}$. In the $\tint=1000 \ \mathrm{hr}$ calculations, confidence contours
of the turning point positions narrowed enough to reveal subtle biases in
their recovered positions, which led to biases in constraints on physical
properties of the IGM as well. These shifts were interpreted to be due to
degeneracies between the signal and the foreground at high frequencies, as
they could be mitigated by using a more sophisticated foreground model or, at
the expense of losing information from turning point D, simply by truncating
the bandpass at 100 MHz. However, even with unbiased constraints on the
turning point positions, biases in the IGM properties persisted, likely because the $\tanh$ is not a perfect
match in shape to the physical model injected into the synthetic dataset. Though we suspect that biases in constraints on the turning points are a persistent feature in datasets employing the tanh, turning point constraints from the $\tint = 100$ hr cases in \citep{Harker2015b} are broad enough to hide such biases. In using the $\tint=100$ hr results, we should then expect to be able to obtain unbiased constraints on the parameters of our physical model in the present work. We also only analyze cases using one and two sky regions, for
which model selection will be more immediately tractable computationally
\citep{Harker2015}.

\begin{figure}[htbp]
\begin{center}
\includegraphics[width=0.48\textwidth]{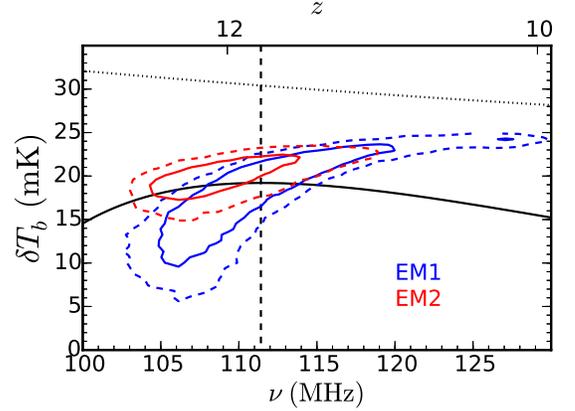}
\caption{Comparison of EM1 and EM2 for turning point D, the point at which
they differ most substantially (see Table \ref{tab:reference_model}). In blue
and red, solid (dashed) contours denote 68\% (95\%) confidence regions for EM1
and EM2, respectively. The dotted black line shows the saturated limit, in
which $\xibar = 0$ and $T_S \approx T_K \gg \Tcmb$, while the dashed vertical
line denotes the position of turning point D in our reference model (solid
black curve). Note that the EM1 error ellipse for turning point D extends to
$\sim 130$ MHz, beyond the edge of the bandpass considered in
\citet{Harker2015b}, though the $2-\sigma$ upper limit for EM2 is within the
assumed band, at $\nu_{\text{D}} \sim 117$ MHz.}
\label{fig:errors_visualized}
\end{center}
\end{figure}

Now, back to the simplest model of \citet{Harker2015b} (EM1 in
Table \ref{tab:reference_model}). This calculation assumed a single sky
region, 100 hours of integration, and a third-order $\log \nu - \log T$
polynomial for the galactic foreground\footnote{It seems likely that in practice a higher order polynomial will be needed to fit out instrumental effects \citep[e.g.,][]{Bernardi2015}. However, here our synthetic datasets contain foregrounds with no structure beyond a polynomial of order three (or four), meaning third and fourth order polynomials can fit the foreground perfectly (by construction).}. \citet{Harker2015b}
investigated the generic case of an idealized instrument (a flat 85\% response
function), though this could easily be modified to enable forecasting for
non-ideal instruments. The foreground and astrophysical signal were
simultaneously fit using the parallel-tempering sampler in the publicly
available \emcee\ code\footnote{\url{http://dan.iel.fm/emcee/current/}}
\citep{ForemanMackey2013}, a \textsc{python} implementation of the
affine-invariant MCMC sampler of \citet{Goodman2010}, from
which constraints on the positions of the turning points followed
straightforwardly. The errors on the turning points are in general not
Gaussian, and are often correlated with one another, though for the purposes of our fitting, we approximate the errors as
1-D independent Gaussians since covariances are likely to depend on the choice of signal parameterization. In addition to EM1, we also investigate the results of a fit using a 4$^\mathrm{th}$ order log-polynomial for the foreground model, which we refer to as EM2. Table \ref{tab:reference_model} summarizes the different
signal extraction models, while Figure \ref{fig:errors_visualized} illustrates
the primary difference between the two extraction models graphically.

%%
% MCMC
%%
\subsection{Parameter Estimation}
With a physical model for the global 21-cm signal (\S\ref{sec:physical_model})
and a set of constraints on the turning point positions (\S\ref{sec:mockobs}),
we then explore the posterior probability distribution function (PDF) for the
model parameters, $\mathbf{\theta}$, given the data, $\mathcal{D}$. That is,
we evaluate Bayes' theorem,
\begin{equation}
    P(\mathbf{\theta} | \mathcal{D}) \propto \mathcal{L}(\mathcal{D} | \mathbf{\theta} ) \mathcal{P}(\mathbf{\theta}) .
\end{equation}
The log-likelihood is given by
\begin{equation}
    \log \mathcal{L} (D | \mathbf{\theta}) \propto -\sum_i \frac{[x (\mathbf{\theta}) - \mu_i]^2}{2 \sigma_i^2}
\end{equation}
where $\mu_i$ is the ``measurement'' with errors $\sigma_i$ (i.e., the values
listed in columns 3 and 4 of Table \ref{tab:reference_model}), and
$x(\mathbf{\theta})$ represents a vector of turning point positions extracted
from the model global 21-cm signal generated with parameters
$\mathbf{\theta}$. This ``two-stage approach'' to fitting the global 21-cm
signal -- the first stage having been conducted by \citet{Harker2015b} -- is
much more tractable computationally than a direct ``one-stage'' fit to a mock
dataset using a physical model. Note that the brightness temperature of
turning point B, $\dTb(\nuB)$, is tightly coupled to its frequency, so we are
effectively only using 5 independent data points in our fits.

\begin{deluxetable}{llccc}
\tabletypesize{\scriptsize}
\tablecaption{Parameter Space Explored}
\tablecolumns{5}
\tablehead{Parameter & Description & Input & Min & Max} 
\startdata
$\Tmin \ (\mathrm{K})$ & Min. virial temp. of star-forming haloes & $10^4$ & $100$ & $10^{5.7}$ \\
$\xi_{\mathrm{LW}}$ &  Ly-$\alpha$ efficiency & 969 & $10$ & $10^6$ \\
$\xi_X$ & X-ray efficiency & 0.02 & $10^{-4}$ & $10^6$ \\
$\xi_{\ioniz}$ & Ionizing efficiency & 40 & $10^{-4}$ & $10^5$
\enddata
\tablecomments{Parameter space explored for results presented in \S\ref{sec:results}. The first two columns indicate the parameter name and a brief description, the third column is the ``true value'' of the parameter in our reference model, while the last two columns indicate the bounds of the priors for each parameter, all of which are assumed to be uninformative, i.e., modeled as uniform distributions between the minimum and maximum allowed values.}
\label{tab:priors}
\end{deluxetable}

To explore this four-dimensional space, we use \textsc{emcee}. We assume
broad, uninformative priors on all parameters (listed in Table
\ref{tab:priors}), but note that our physical model implicitly imposes three
additional constraints on the astrophysical signal:
\begin{enumerate}
    \item We neglect exotic heat sources at high-$z$, which confines turning point B to a narrow ``track'' at $\nu \lesssim 50$ MHz.
    \item We assume that the Universe cannot cool faster than the Hubble expansion, which sets a redshift-dependent lower limit on the strength of the absorption signal (lower dotted curve in all panels of Figure \ref{fig:param_study}).
    \item We assume the mean density of the IGM we observe is the universal mean density, i.e., it has $\delta = 0$, which prevents the signal from exceeding the ``saturated limit,'' in which $T_S \gg \Tcmb$ and $x_i = 0$ (upper dotted curve in all panels of Figure \ref{fig:param_study}).
\end{enumerate}
Our code could be generalized to accommodate exotic heating models, though
this is beyond the scope of this paper. Bullets 2 and 3 above
are manifestly true for gas at the cosmic mean density (via Equation
\ref{eq:dTb}), though imaging campaigns will likely see patches of IGM whose
brightness temperatures exceed (in absolute value) these limits, owing to
over-densities $\delta > 0$.

For all calculations presented in this work, we use 384 \textit{walkers}, each
of which take a 150 step burn-in, at which point they are re-initialized in a
tight ball centered on the region of highest likelihood identified during the
burn-in. We then run for 150 steps more (per walker), resulting in MCMC chains
with 57,600 links. The mean acceptance fraction, i.e., the number of proposed
steps that are actually taken during our MCMC runs, is $\sim 0.3$. The runs
are well-converged, as we see no qualitative differences in the posterior
distributions when we compare the last two 10,000 element subsets of the full
chain.

%%%
%% RESULTS
%%%
\section{Results} \label{sec:results}
Each MCMC fit yields 57,600 samples of the posterior distribution, which is a
4-dimensional distribution in $\{\Tmin, \xiLW, \xiXR, \xiUV \}$ space.
However, we also analyze each realization of the global 21-cm signal
on-the-fly as the MCMC runs, saving IGM quantities of interest every $\Delta z
= 0.1$ between $5 \leq z \leq 35$, as well as at the turning points. To build
upon the analytical arguments presented in Paper I, which provided a basis for
interpreting the turning points in terms of IGM properties, we start with an
analysis of the inferred IGM properties at the turning points in
\S\ref{sec:IGMconstraints}, deferring a full analysis of the IGM history to
future work. Readers interested only in the constraints on our four-parameter
model can proceed directly to \S\ref{sec:4parconstraints}.

%%
% IGM
%%
\subsection{Constraints on the Intergalactic Medium} \label{sec:IGMconstraints}
\begin{figure}[htbp]
\begin{center}
\includegraphics[width=0.48\textwidth]{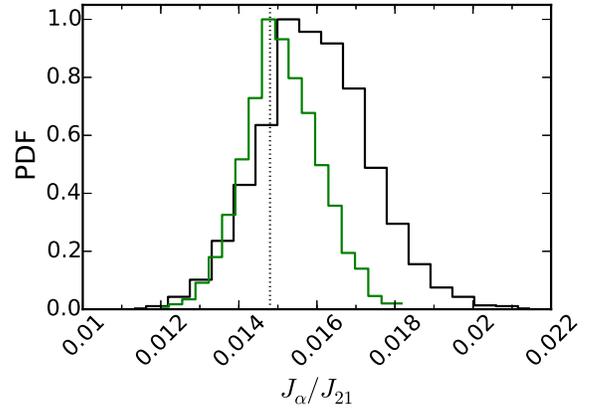}
\caption{Constraints on the background $\Lya$ intensity at the redshift of turning point B, in units of $J_{21}=10^{-21} \intensityunitsenergy$. Dotted vertical line shows the input value, which occurs at $z=29$ in our reference model. The black histogram is the constraint obtained if using EM1, while the analogous constraint for EM2 is shown in green. }
\label{fig:igm_B}
\end{center}
\end{figure}

\begin{figure}[htbp]
\begin{center}
\includegraphics[width=0.48\textwidth]{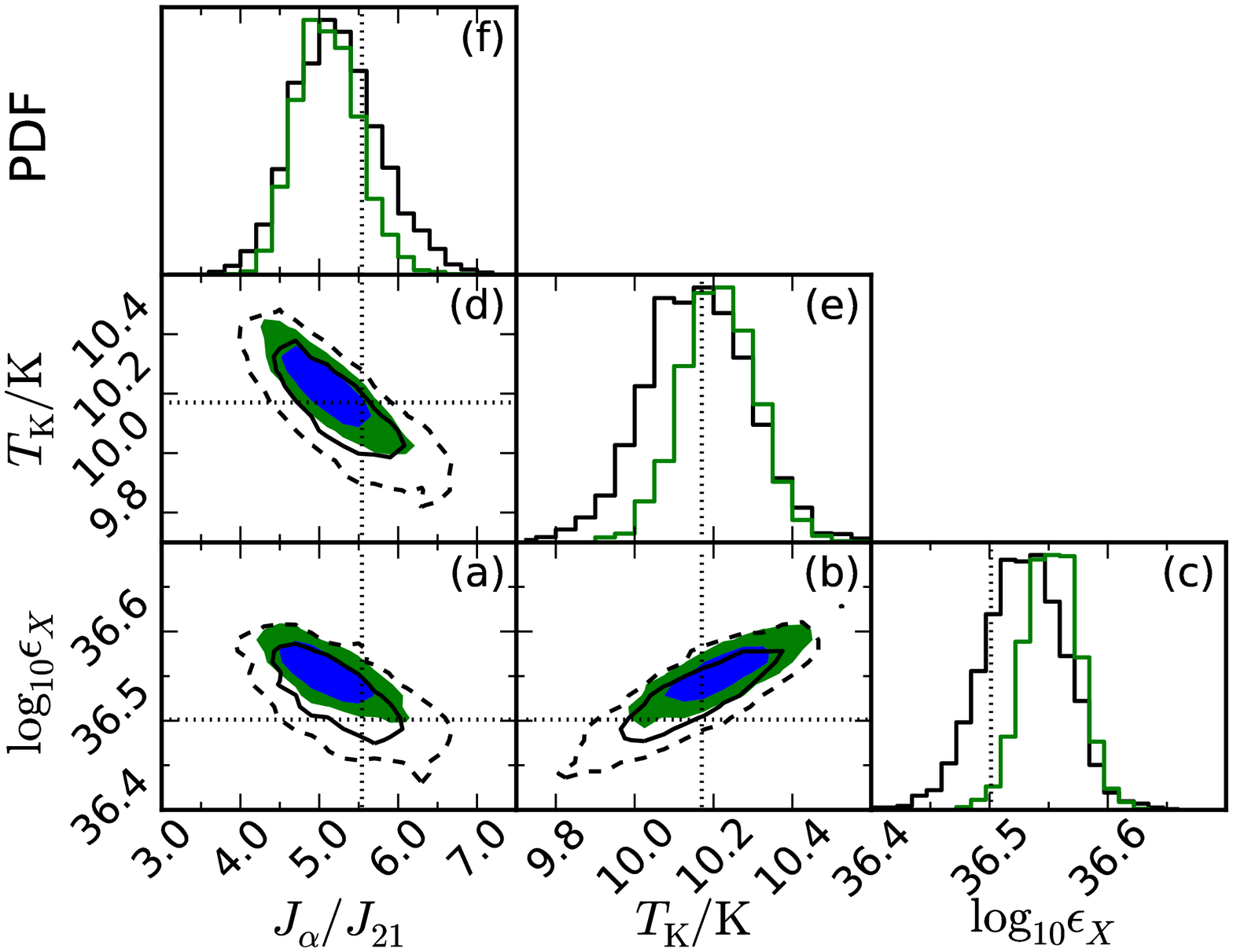}
\caption{Constraints on the $\Lya$ background intensity, IGM temperature, heating rate density at the redshift of turning point C. The heating rate density, $\eX$, is expressed in units of $\coheatingdensity$, while $\Jalpha$ is once again expressed in units of $J_{21}=10^{-21} \intensityunitsenergy$. Dotted vertical lines show the input values, which occur at $z=19$ in the reference model. Open contours are those obtained with EM1 (68\% and 95\% confidence regions in solid and dashed curves, respectively), while filled contours are the constraints obtained by EM2 (68\% and 95\% confidence regions in blue and green, respectively). The color-scheme along the diagonal is the same as in Figure \ref{fig:igm_B}, with EM1 (EM2) curves in black (green).}
\label{fig:igm_C}
\end{center}
\end{figure}

\begin{figure*}[htbp]
\begin{center}
\includegraphics[width=0.98\textwidth]{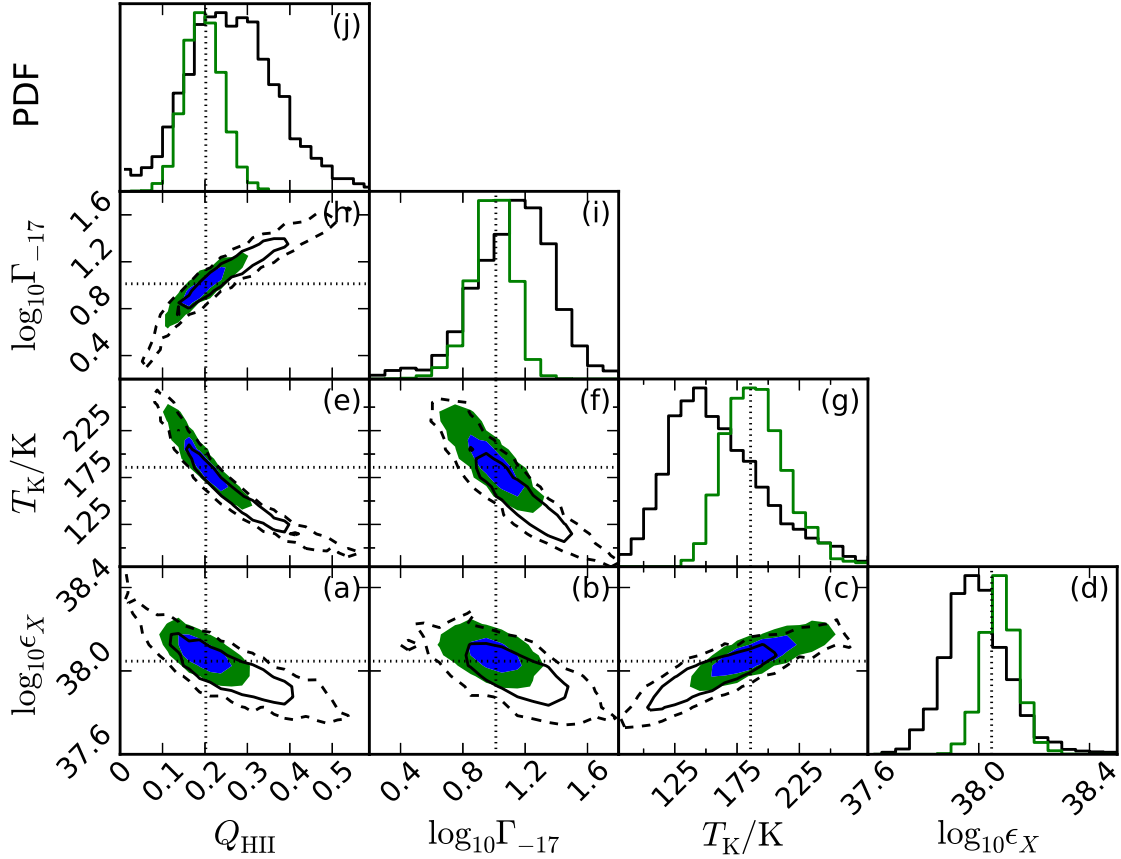}
\caption{Constraints on the volume-filling factor of ionized gas, volume-averaged ionization rate, IGM temperature, and heating rate density at the redshift of turning point D, which occurs at $z=11.75$ in our reference model. Open contours are 68\% (solid) and 95\% (dashed) confidence regions for EM1, while filled contours show the results from EM2, with 68\% and 95\% confidence regions shown in blue and green, respectively. Input values are denoted by black dotted lines in each panel.}
\label{fig:igm_D}
\end{center}
\end{figure*}

We begin by showing our mock constraints on properties of the IGM at the redshifts of turning points B, C, and D in Figures \ref{fig:igm_B}, \ref{fig:igm_C}, and \ref{fig:igm_D}, respectively\footnote{Note that our choice to derive constraints on the IGM parameters at the turning points, rather than at a fixed series of redshifts, is in part responsible for some of the behavior in Figures \ref{fig:igm_C} and \ref{fig:igm_D}. This seemed a natural choice given that we fit our model to the turning points alone. Constraints at an arbitrary redshift could be extracted in future studies, for example to compare to independent measurements that do not coincide with the redshifts of the turning points.}. 
 
Because turning point B primarily probes the $\Lya$ background, we focus only on its ability to constrain $\Jalpha$ in Figure \ref{fig:igm_B}. The input value is recovered to $1-\sigma$, with relatively tight error-bars limiting $\Jalpha$ to within a factor of 2. Use of EM2 has little effect on this constraint as its main improvement over EM1 is at frequencies $\nu \gtrsim 100$ MHz.

Figure \ref{fig:igm_C} shows constraints on the $\Lya$ background and thermal
history at the redshift of turning point C. In the $\sim 90$ Myr separating
turning points B and C, the $\Lya$ background intensity, $\Jalpha$, has
risen by a factor of $\sim 350$, though we still constrain its value to within
a factor of $\sim 2$ (panel f). The IGM temperature is limited to $9 \lesssim
T_K / \mathrm{K} \lesssim 11$, and would otherwise be $\sim 7.4$ K at this
redshift in the absence of heat sources. There are noticeable degeneracies in
the 2-D PDFs, which are not necessarily obvious intuitively.

Let us first focus on the anti-correlations in the $\Jalpha$-$\eX$ and $\Jalpha-T_K$ planes (panels a and d in Figure \ref{fig:igm_C}). For this exercise -- and those that follow -- it will be useful to consider slight excursions away from our reference model. We can see from the lower right panel of Figure \ref{fig:param_study} that a small increase in $\xiLW$ will shift turning point B to slightly higher redshifts (lower frequencies) holding all other parameters fixed. Turning point C will also occur earlier than in our reference model (since a stronger $\Lya$ background can couple $T_S$ to $T_K$ more rapidly ) and be deeper, since there has been less time for X-rays to heat the IGM, leading to increased contrast between the IGM and the CMB. Panels (a) and (d) in Figure \ref{fig:igm_C} now make sense: the anti-correlations in the $\Jalpha$-$\eX$ and $\Jalpha-T_K$ planes arise because measurement errors permit slight excursions away from the reference model, which if achieved through enhancements to $\xiLW$, shift turning points B and C to slightly earlier -- and thus cooler -- times. 

One could also counteract a mild increase in $\xiLW$ with a corresponding increase in $\xiXR$, which enhances heating and thus leads to shallower absorption troughs. However, increasing $\xiXR$ shifts turning point C to shallower depths \textit{and} lower frequencies, thus exacerbating the leftward shift caused by larger values of $\xiLW$. As a result, $\Tmin$ would also need to be increased in order to delay the onset of Wouthuysen-Field coupling and heating. Indeed, we will find this series of positive correlations among the physical parameters of our model in the \S\ref{sec:4parconstraints}.

Before moving on to the IGM constraints associated with turning point D, we
note that the correlation between $T_K$ and $\eX$ (panel b) is simply because
$T_K \propto \int \eX dz$, and $\eX$ is monotonic. Also, apparently the
improvement at the highest frequencies offered by EM2 also acts to slightly
bias constraints on $\Jalpha$ and $\eX$ relative to their input values.
Referring back to Figure \ref{fig:errors_visualized}, we do see a slight bias
in the EM2 PDF for turning point D toward larger amplitude, which would
require more rapid heating at earlier times. In fact, this is precisely the
sense of the bias we see in Figure \ref{fig:igm_C}: slightly larger values of
$\eX$ at turning point C, and a corresponding downward shift in $\Jalpha$ as
described above.

And finally, Figure \ref{fig:igm_D} shows constraints on the ionization and
thermal histories at the redshift of turning point D, which occurs at $z =
11.75$ in our input model. The behavior here is complex, as the signal is not
yet saturated (i.e., $T_S \gg \Tcmb$ is a poor approximation) and the mean
ionized fraction is non-zero (i.e., $\QHII \sim 0.2$). This means the global
21-cm signal depends on both the ionization history and the thermal history,
which we may parameterize in terms of the volume filling factor of ionized
gas, $\QHII$, the IGM temperature, $T_K$, and their
time-derivatives\footnote{Although we use the symbol $\Gamma$, we caution that
our values should not be compared to extrapolations of constraints on
$\Gamma_{\HI}$ from the $\Lya$ forest at $z \lesssim 6$. The latter is a probe
of the meta-galactic ionizing background (i.e., large-scale backgrounds),
whereas our values of $\Gamma$ probe the growth rate of ionized regions, and
thus should be considered a probe of radiation fields \textit{near} galaxies.
A more detailed cosmological radiative transfer treatment could in principle
reconcile the two tracers of ionizing sources.} $\Gamma_{\HI}$ and $\eX$. We
may, however, neglect the $\Lya$ history at this stage, since $T_S \approx
T_K$ is accurate to high precision, rendering any constraints on $\Jalpha$
completely parameterization-dependent (i.e., $\Jalpha$ can be anything, so
long as it is large enough to drive $T_S \rightarrow T_K$).

It is once again useful to consider excursions away from the reference model. At fixed thermal history, a slight increase in $\xiUV$ will act to decrease the amplitude of turning point D and shift it to slightly higher redshift. With less time to heat the IGM between turning points C and D, the IGM is cooler at the redshift of turning point D in this scenario and as a result, the emission signal is weaker than that of our reference model. This line of reasoning explains the anti-correlations between the ionization and thermal history parameters in Figure \ref{fig:igm_D}. As in Figure \ref{fig:igm_C}, positive correlations occur by construction, since state quantities like $\QHII$ and $T_K$ are just integrals of $\Gamma_{\HI}$ and $\eX$, which are both monotonically increasing with decreasing redshift.

The advantages of EM2 over EM1 are also clear in Figure \ref{fig:igm_D}. This
improvement occurs because EM1 does not detect turning point D with
significance away from the saturated limit or within the assumed band ($\nu
\leq 120$ MHz), whereas the EM2 fit does both at the $> 2-\sigma$ level.
Perhaps most notably, this leads to a strong detection of the early stages of
reionization ($0.12 \leq \QHII \leq 0.29$ at $2-\sigma$; green PDF in panel j).

Lastly, we note that although the amplitude of the signal is set by $\xibar$,
a volume-averaged ionized fraction, we only show constraints on $\QHII$, as
$x_e$ never reaches values above $\sim 10^{-2}$ at $z \gtrsim 10$ in any of
our calculations. As a result, it has a negligible impact on the ionization
history. However, even mild ionization of the bulk IGM enhances the efficiency
of heating rather substantially since the fraction of photo-electron energy
deposited as heat (as opposed to ionization or excitation) is a strong
function of the electron density \citep[e.g.][]{Shull1985,Furlanetto2010},
which means the value of $x_e$ can have a considerable effect on the thermal
history of the IGM. Our choice of a mean X-ray photon energy of $h\nu_X = 0.5$
keV, in lieu of a detailed solution to the radiative transfer equation, drives
this result. More detailed calculations that solve the RTE
\citep[e.g.][]{Mirocha2014} could enable scenarios in which the bulk IGM is
ionized substantially prior to the overlap phase of reionization, which could
have interesting observational signatures. We defer a more detailed treatment
of this effect to future work.

%%
% REFERENCE MODEL
%%
\subsection{Constraints on the Physical Model} \label{sec:4parconstraints}
\begin{figure*}[htbp]
\begin{center}
\includegraphics[width=0.98\textwidth]{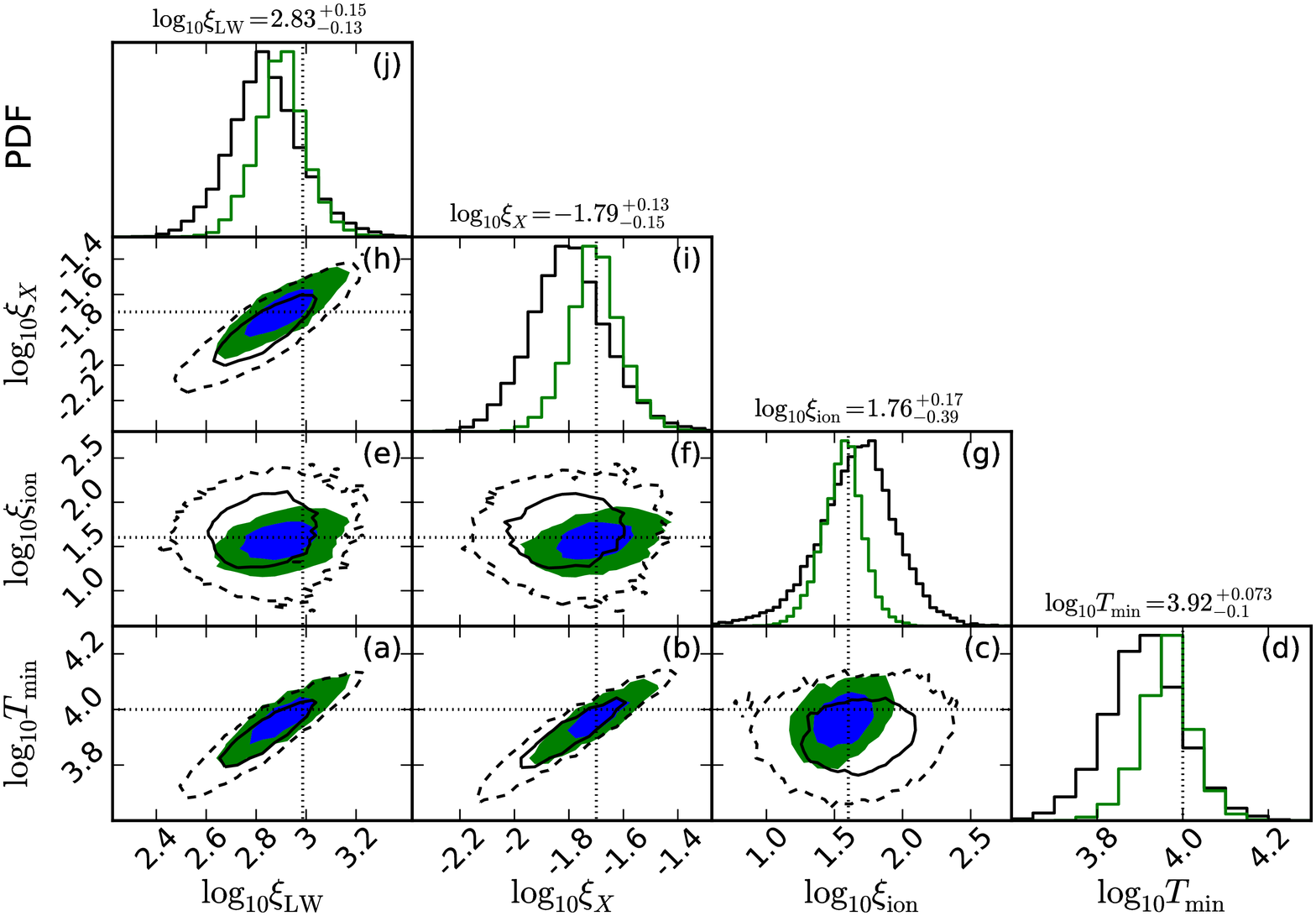}
\caption{Constraints on our 4-parameter reference model. Filled contours in the interior panels are 2-D marginalized posterior PDFs with 68\% confidence intervals shaded blue and 95\% confidence regions in green. Panels along the diagonal are 1-D marginalized posterior PDFs for each input parameter, with 1-$\sigma$ asymmetric error-bars quoted, as computed via the marginalized cumulative distribution functions. Dotted lines denote the input values of our reference model (Table \ref{tab:reference_model}). Bins of width 0.05 dex are used in each panel. Annotated best-fit values and error bars along the diagonal are those associated with EM2.}
\label{fig:triangle_1_100_4par}
\end{center}
\end{figure*}

\begin{figure*}[htbp]
\begin{center}
\setlength{\unitlength}{1cm}
\hspace*{-0.4cm}\resizebox{9.2cm}{!}{\includegraphics[width=0.45\linewidth]{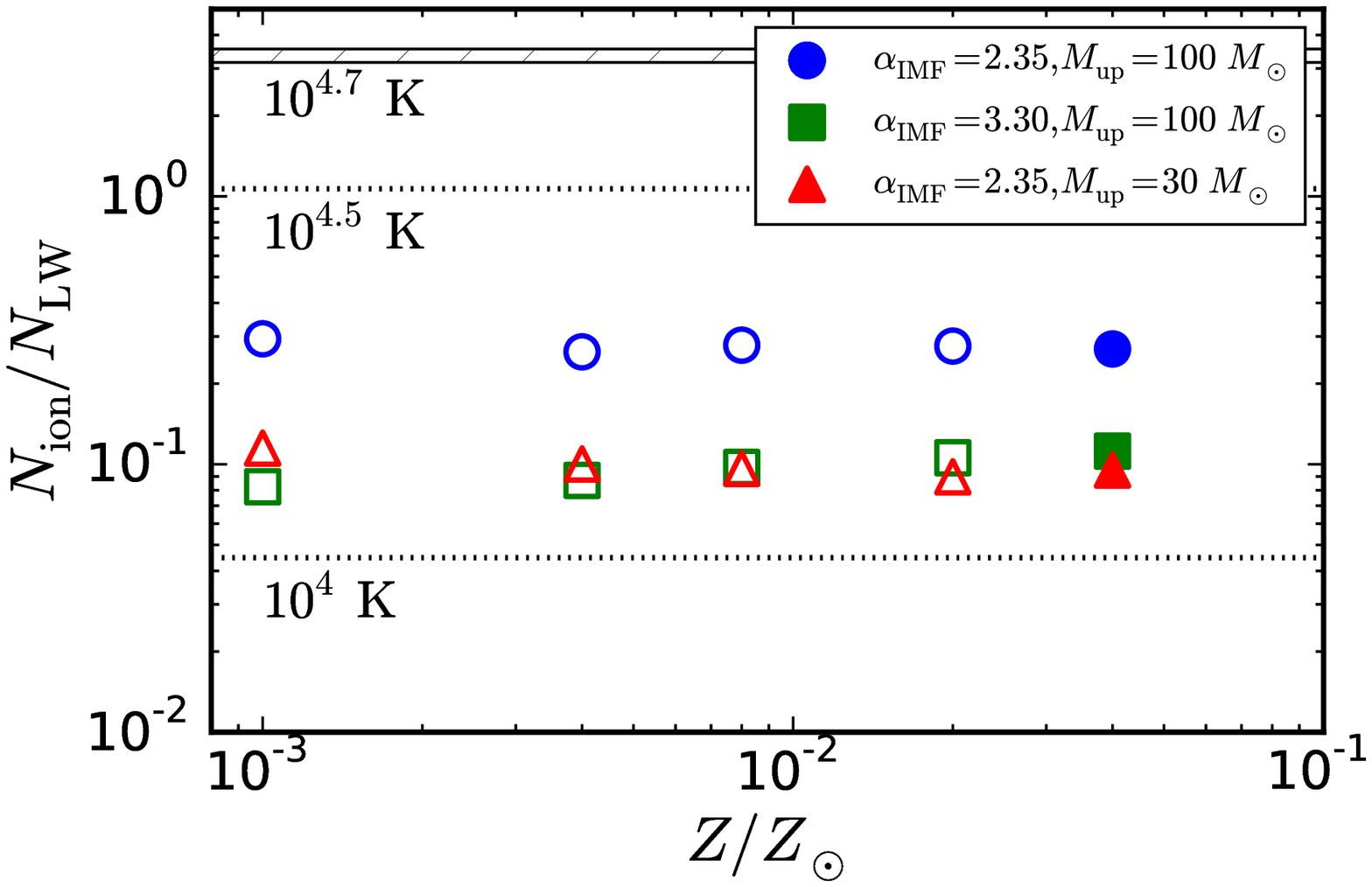}}
\hspace*{-0.4cm}\resizebox{9.2cm}{!}{\includegraphics[width=0.45\linewidth]{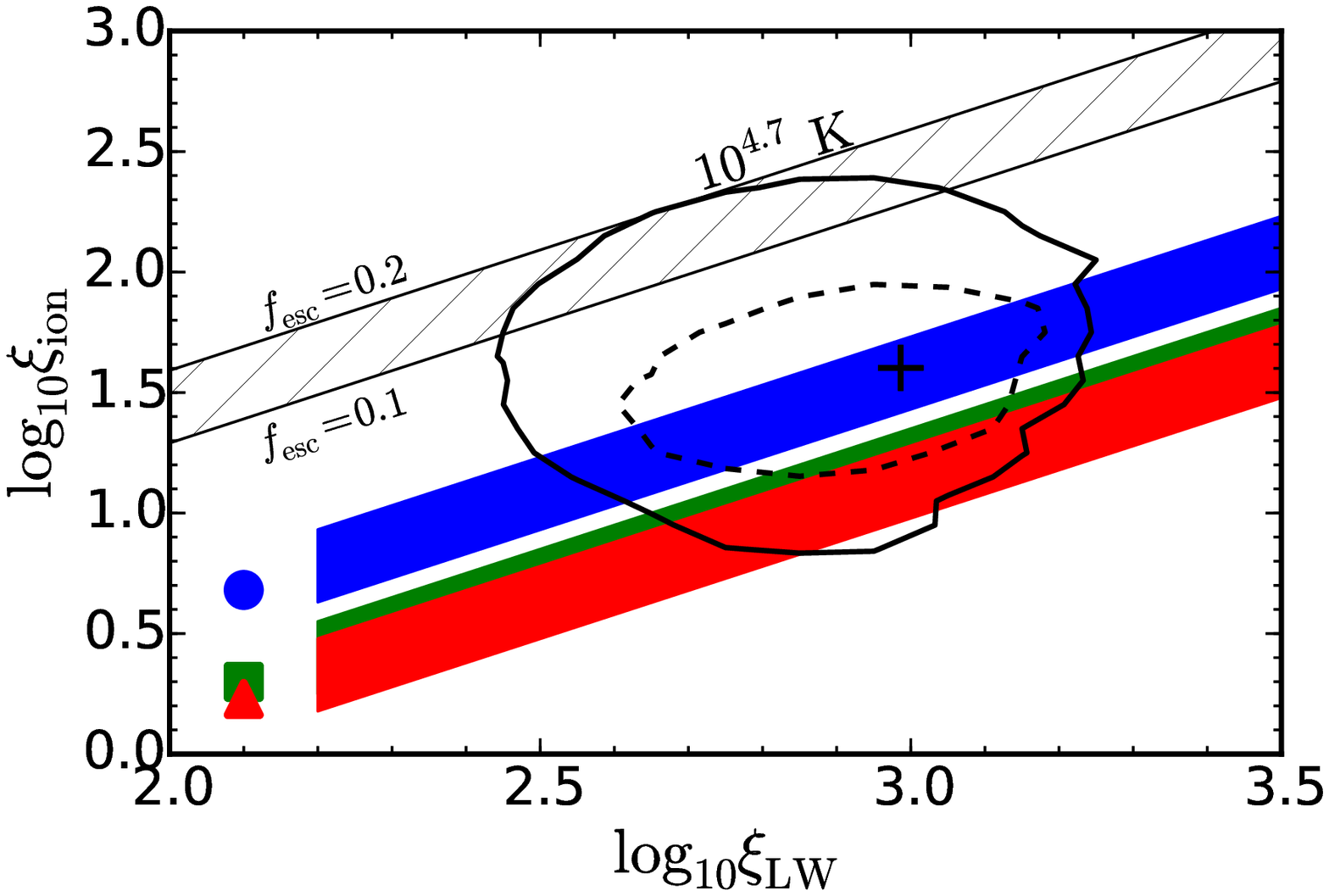}}
\caption{\textit{Left}: Ratio of yields in the ionizing ($h\nu > 13.6$ eV) and LW ($10.2 \leq h\nu / \mathrm{eV} \leq 13.6$) bands per stellar baryon as a function of metallicity and stellar IMF. Symbols represent model SEDs generated with \textsc{starburst99} \citep[those shown in Figures 1, 3, and 5 of][]{Leitherer1999}, while the horizontal lines show the values one obtains for pure blackbodies at $10,000$, $30,000$, and $50,000$ K from bottom to top. The filled symbols are investigated in more detail in the right panel. \textit{Right:} Constraints on the stellar population and the escape fraction of ionizing radiation. The solid contour is the 2-$\sigma$ constraint on our reference model, i.e., identical to the green area of panel (e) in Figure \ref{fig:triangle_1_100_4par}, while  dashed contours correspond to turning point constraints from EM2 (see Table \ref{tab:reference_model}), which has a tighter constraint on the emission maximum (turning point D). The blue, green, and red bands have the same value of $\Nion/\Nlw$ as the filled plot symbols in the left-hand panel, while the cross-hatched band instead adopts a pure $50,000$ K blackbody spectrum for the stellar population. The width of each band corresponds to a factor of 2 change in the escape fraction, $0.1 \leq \fesc \leq 0.2$.}
\label{fig:stellar_pop}
\end{center}
\end{figure*}

\begin{figure}[htbp]
\begin{center}
\includegraphics[width=0.48\textwidth]{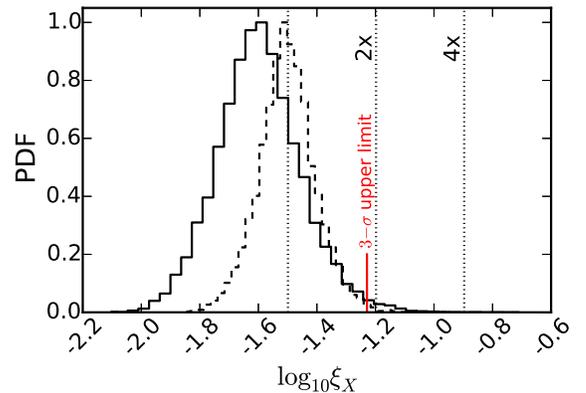}
\caption{Constraints on sources of X-rays. The solid and dashed histograms are identical to the black and green PDFs shown in panel (i) of Figure \ref{fig:triangle_1_100_4par}, respectively, while dotted vertical lines denote values representative of 2x and 4x enhancements to the $L_X$-SFR relation, holding the star-formation efficiency constant at $\fstar=0.1$. For reference, we show the $3-\sigma$ upper limit from extraction EM2 in red. Even if $\fstar=0.01$, we limit $f_X < 20$.}
\label{fig:bh_pop}
\end{center}
\end{figure}

Our main results are illustrated in Figures \ref{fig:triangle_1_100_4par},
\ref{fig:stellar_pop}, and \ref{fig:bh_pop}, which analyze the full 4-D
constraints on our reference model and the implications for UV and X-ray sources, respectively. In this section, we'll examine each in turn.

It is perhaps most intuitive to begin with the panels along the diagonal of
Figure \ref{fig:triangle_1_100_4par}, which show the marginalized 1-D
constraints on the parameters of our reference model. As predicted, given its
broad-band influence on the signal, $\Tmin$ (panel d) is most tightly
constrained, with $1\sigma$ error bars of order $\sim 0.05$ dex. Therefore, an
idealized instrument observing a single sky region for 100 hours can rule out
star-formation in molecular halos (onto which gas collapses more slowly; see
Figure \ref{fig:fcoll_2panel}), at least at levels sufficient to affect all
three turning points. Errors on $\xi_X$ and $\xi_{\mathrm{LW}}$ are comparable
(panels i and j), though the positive error-bars are larger at $\sim 0.1$ dex.
The errors on $\xi_{\ioniz}$ are more asymmetric, at $+0.1/-0.2$ dex (panel
g).

Strong degeneracies are also apparent, particularly in panels (a), (b), and
(h), which show 2-D constraints in the $\Tmin$--$\xi_{\mathrm{LW}}$,
$\Tmin$--$\xi_X$, and $\xi_{\mathrm{LW}}$--$\xi_X$ planes, respectively. The
first two are straightforward to understand. An increase in
$\xi_{\mathrm{LW}}$ drives an enhancement in $\Lya$ production \textit{per
unit star-formation}, which can be counteracted by a reduction in the
star-formation rate \textit{density}. In our modeling framework, a reduction
in the SFRD is achieved by increasing $\Tmin$, confining star formation to
more massive and thus more rare halos. If $\fstar$ were allowed to vary, it
too could limit the SFRD, though the change would be systematic, whereas
varying $\Tmin$ affects both the normalization and the redshift evolution. The
same line of reasoning explains the relationship between $\Tmin$ and $\xi_X$.

The $\xi_{\mathrm{LW}}$--$\xi_X$ degeneracy is slightly more complex. An
increase in $\xi_{\mathrm{LW}}$ seeds a stronger $\Lya$ background (holding
$\Tmin$ fixed), which in turn shifts turning point B to lower frequencies (see
the lower right panel of Figure \ref{fig:param_study}), which measurement
error permits to some degree. This will result in a deeper (and earlier)
absorption trough \textit{unless} $\xi_X$ is increased, causing a shallower
trough (see upper right panel of Figure \ref{fig:param_study}). Once again,
measurement error sets the degree to which an increase in $\xi_X$ can
compensate for an increase in $\xi_{\mathrm{LW}}$. As discussed in
\S\ref{sec:IGMconstraints}, slight excursions in $\xiXR$ cannot completely
correct for changes in $\xiLW$, and will require changes in $\Tmin$ as well,
especially if the measurement errors are small. In the limit of very large
error-bars, however, confidence contours would not close and instead we would
have large ``bands'' through parameter space, signifying an insurmountable
degeneracy between two parameters. Our results indicate that observations of a
single sky region for 100 hours, albeit with an idealized instrument, are
precise enough to close these contours, and recover all input values to within
$1-\sigma$ confidence. We will revisit this claim in \S\ref{sec:discussion}.

At this stage it may seem like we have just traded constraints on one set of
phenomenological parameters (the parameters of the tanh model; Equation \ref{eq:tanhdef}) for another ($\Tmin$ and
the $\xi$'s). However, if we assume that $\xi_{\mathrm{LW}}$ and
$\xi_{\ioniz}$ probe the same stellar population, their ratio is independent
of the star-formation history (which is set by $\fstar$ and $\Tmin$, which we assume are time-independent), and thus constrains the spectral energy distribution (SED) of galaxies modulo a factor of the escape fraction\footnote{We assume the escape fraction of LW photons is 100\%, though in reality this is only likely to be true in the smallest halos \citep[e.g.,][]{Kitayama2004}. For simplicity, we neglect this complication and defer a more thorough treatment to future work.}, i.e. (following Equations \ref{eq:xiLW} and \ref{eq:xiUV}),
\begin{equation}
    \frac{\xi_{\ioniz}}{\xi_{\mathrm{LW}}} = \frac{N_{\ioniz}}{\Nlw} \fesc . \label{eq:stellar_pop}
\end{equation}
To compute $N_{\ioniz}/\Nlw$, we take model spectral energy distributions
directly from \citet{Leitherer1999}. We focus on those assuming an
instantaneous burst of star-formation with nebular emission included (their
Figures 1, 3, and 5), and find the cumulative number of photons emitted in the
LW and hydrogen-ionizing bands, which typically plateaus around
$\sim 20$ Myr after the initial burst. The results, as a function of
metallicity and stellar initial mass function (IMF), are shown in the left
panel of Figure \ref{fig:stellar_pop}. While the values of $N_{\ioniz}$ and
$\Nlw$ vary by factors of $\sim 2$ over the metallicity range $0.001 \leq Z /
Z_{\odot} \leq 0.04$, their \textit{ratio} changes by only $\sim 0.05$ over
this same interval in metallicity. The more important dependence is on the
stellar IMF: a standard Salpeter IMF, with $\alpha_{\mathrm{IMF}} = 2.35$ and
an upper mass cutoff of $M_{\mathrm{cut}} = 100 \ \Msun$, yields $0.25
\lesssim N_{\ioniz}/\Nlw \lesssim 0.3$ for all $10^{-3} \leq Z / Z_{\odot}
\leq 0.04$ (blue circles in the left panel of Figure \ref{fig:stellar_pop}),
whereas mass functions with fewer massive stars, whether that be achieved with
steeper power-law indices ($\alpha_{\mathrm{IMF}} = 3.3$; green squares in
Figure \ref{fig:stellar_pop}) or by reducing the upper cutoff
($M_{\mathrm{cut}} = 30 \ \Msun$; red triangles in Figure
\ref{fig:stellar_pop}), yield $0.07 \lesssim N_{\ioniz}/\Nlw \lesssim 0.12$.

In the right panel of Figure \ref{fig:stellar_pop}, we compare our constraints
in the $\xiLW-\xiUV$ plane with the $Z=0.04 \ Z_{\odot}$ stellar population
models described above. The red, green, and blue bands in the right panel
correspond to the stellar population models denoted by filled points of the
same color in the left panel. We also show the case of a pure $50,000$ K
blackbody spectrum in the cross-hatched region. The width of each band
corresponds to a factor of two change in the escape fraction, $0.1 \leq \fesc
\leq 0.2$.

Our mock constraints on $\xiLW/\xiUV$ given 100 hours of integration on a
single sky region (EM1) can only rule out rather extreme cases. For example,
this scenario rules out the $50,000$ K toy stellar population with $\fesc
\gtrsim 0.2$ at one extreme, and bottom-heavy IMFs with escape fractions below
$\fesc \lesssim 0.1$ at the other extreme. A stronger detection of turning
point D, achieved by EM2, tightens these constraints considerably. The pure
$50,000$ K stellar population would require $\fesc \lesssim 0.01$, while a
stellar population with an over-abundance of lower mass stars would require
$\fesc \gtrsim 0.2$. Note that the surface temperatures of PopIII stars are
expected to be $\sim 10^5$ K, which only strengthens our limits quoted for the
$50,000$ K population. Our reference model assumes a typical PopII stellar
population, so it is reassuring to see that our constraints coincide with the
blue diagonal band, which represents a standard Salpeter IMF.

Synthesis models for black hole populations are growing in maturity, though
still only loosely constrained by observations, especially at low
metallicities \citep[e.g.,][]{Belczynski2008}. An immediate interpretation of
$\xi_X$ will thus be very challenging barring progress on this front in the
coming years, given that we cannot eliminate the degeneracy with the SFRD as
we did previously by looking at the ratio $\xiLW/\xiUV$, rather than either
quantity independently. For simplicity, we assume an $\alpha = 1.5$ power-law
spectrum above 0.2 keV consistent with the findings of \citet{Mineo2012a}, and
$\fstar=0.1$. The 1-D marginalized PDFs for $\xi_X$ for EM1 and EM2 are shown
in Figure \ref{fig:bh_pop}. Factor of $\sim$ a few enhancements are allowed
out to $z \lesssim 4$ \citep{Dijkstra2012,BasuZych2013}, though the redshifts
probed by the global 21-cm signal are far beyond the reach of the techniques
used to establish such limits (the cosmic X-ray background and image stacking,
respectively). All signal extraction models considered here rule out factor of
2 enhancements to $f_X$ at the $\sim 3 \sigma$ level assuming $\fstar=0.1$. We
will revisit this type of constraint in \S\ref{sec:discussion}.

Our reference model is seemingly inconsistent with star formation in molecular
halos and a stellar IMF that yields more high-mass X-ray binaries than average
(per unit star formation). This does not rule out star formation in molecular
halos or a top-heavy IMF, \textit{it just rules out such sources as important
drivers of the turning points}. If we assume that PopIII stars have $\Nlw =
4800$, a SFRD of $\approx 3 \times 10^{-7} \ \sfrd$ would be required to match
the constraint on $\Jalpha$ provided by turning point B (following
Eq. 17 of Paper I), which corresponds to $\fstar \approx 0.1$ in
$\Tmin=300$ K halos. Such a population would have to die out rapidly in order
for turning point C to be unaffected. Put another way, if PopIII stars do form
relatively efficiently at $z \sim 30$, and continue to do so for more than
$\sim 100$ Myr, we should expect the position of turning point C to change
(relative to our reference model) due to a stronger $\Lya$ background and
potentially a stronger X-ray background, depending on the properties of PopIII
remnants.

%%%
%% DISCUSSION
%%%
\section{Discussion} \label{sec:discussion}
Our results suggest that simultaneous fits to the three spectral turning
points of the global 21-cm signal can yield powerful constraints on the
properties of the Universe's first galaxies. A simple 4-parameter model can be
constrained quite well in only 100 hours of integration on a single sky
region, provided an ideal instrument. The $\xi$
parameters place interesting constraints on the properties of the first
generations of stars and black holes, while constraints on the characteristic
redshift-dependent mass of star-forming galaxies follows immediately from
constraints on $\Tmin$. In this section, we discuss these findings within a
broader context, focusing in particular on how our results depend on the
assumed measurement (\S\ref{sec:band_effects}) and model
(\S\ref{sec:assumptions}), and how our fitting procedure might
be improved to maximize the return from ongoing and near-future observing
campaigns (\S\ref{sec:selection}).

%%
% INFO STUDY
%%
\subsection{Are all three points necessary?} \label{sec:band_effects}
Our forecasts have so far assumed that all three spectral features in the $40
\lesssim \nu / \ \mathrm{MHz} \lesssim 120$ window are detected and
characterized reasonably well, apart from the EM1 detection of turning point D
which was only marginal. Given practical limitations in constructing an
instrument with a broad-band response, the ionospheric challenges at low
frequencies, and a weak emission feature potentially plagued by terrestrial
radio frequency interference (RFI), it is worth asking: must we detect all
three features at once to constrain even the simplest of galaxy formation
models?

\begin{figure*}[htbp]
\begin{center}
\includegraphics[width=0.98\textwidth]{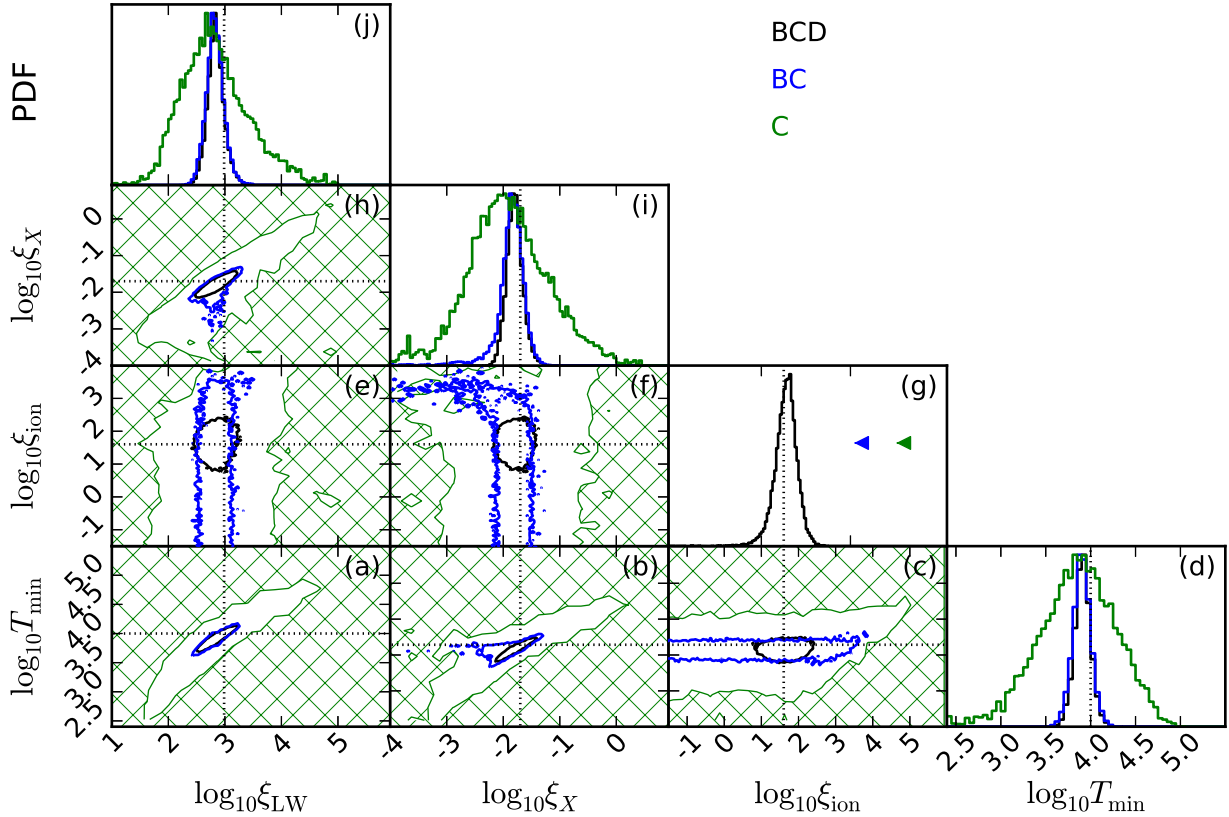}
\caption{Constraints on the 4-parameter model, assuming the emission maximum is not detected. Solid black contours are the 95\% confidence intervals of our reference fit to all three features. Blue contours are 95\% confidence regions obtained when only turning points B and C are used in the fit, while the green hatched regions show ares of parameter space excluded at 95\% confidence if only turning point C used in the fit. When only upper or lower limits are available, we denote them with arrows in the marginalized 1-D PDFs.}
\label{fig:BCDinfo_lowfreq}
\end{center}
\end{figure*}

\begin{figure*}[htbp]
\begin{center}
\includegraphics[width=0.98\textwidth]{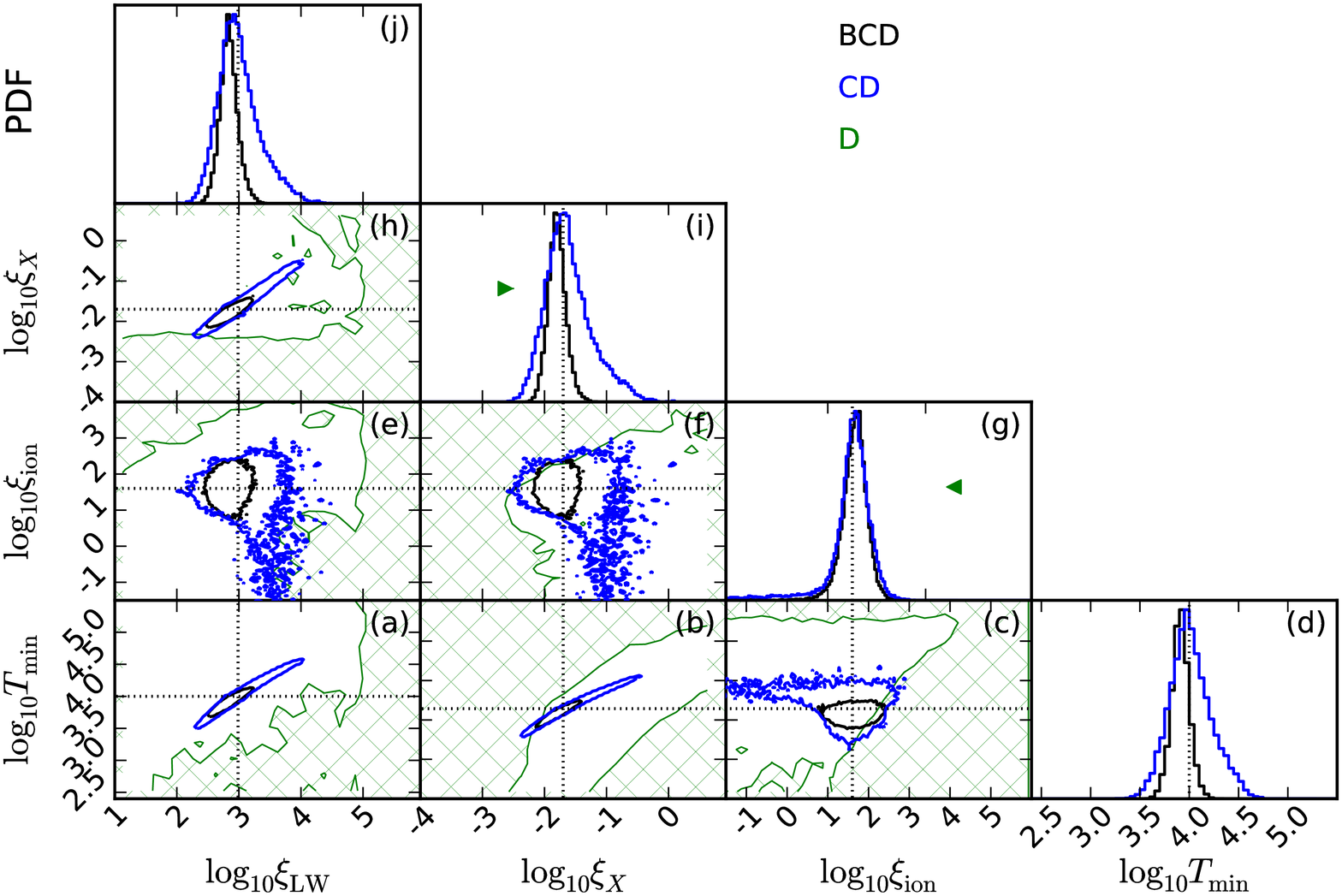}
\caption{Constraints on the 4-parameter model, assuming turning point B is not detected. Solid black contours are the 95\% confidence intervals of our reference fit to all three features. Blue contours are 95\% confidence regions obtained when only turning points C and D are used in the fit, while the green hatched regions show ares of parameter space excluded at 95\% confidence if only turning point D used in the fit. When only upper or lower limits are available, we denote them with arrows in the marginalized 1-D PDFs.}
\label{fig:BCDinfo_hifreq}
\end{center}
\end{figure*}

Figures \ref{fig:BCDinfo_lowfreq} and \ref{fig:BCDinfo_hifreq} show the
constraints on our 4-parameter model assuming only a subset of the turning
points are detected. We consider all possible cases, except a scenario in
which only turning point B is detected, as it seems unlikely that one could
recover this feature from the foreground without help from neighboring
spectral structure, given its amplitude of $\lesssim 5$ mK. Note that the
black contours in each plot are identical to the 95\% confidence regions in
Figure \ref{fig:triangle_1_100_4par}, though the x and y scales of each
individual panel here are much broader than those in Figure
\ref{fig:triangle_1_100_4par} due to the reduced quality of constraints. Blue contours
denote fits including two turning points, while green cross-hatched regions
correspond to fits including only a single turning point. Because the PDFs for
the one- and two-point fits are broad, they tend to become noisy. This
behavior is expected: by design, walkers spend less time in low-likelihood
regions. If those regions of parameter space are large (which they are for the
one- and two-point fits), it will take a long time to properly explore them.

In Figure \ref{fig:BCDinfo_lowfreq}, we focus on the case in which the
emission maximum, turning point D, is not used in the fit. In the most
optimistic case, both turning points B and C are still detected, and give rise
to the constraints shown in blue. As expected, constraints on $\xiUV$ are
virtually nil except for a weak upper limit (panel g). However, constraints on
$\xiLW$, $\xi_X$, and $\Tmin$ remain largely intact. The subtle detours away
from the black contours in panels b, f, and h toward small values of $\xi_X$
are real: they indicate scenarios in which heating is negligible and turning
point C is induced by ionization (see \S3.2.2 in Paper I). Such models would
likely lead to an early end to the EoR and thus a large value of the Thomson
optical depth, $\tau_e$, though without a detection of turning point D or a
prior on $\tau_e$ such scenarios remain allowed. In a more pessimistic
scenario in which only the absorption minimum, turning point C, is detected,
$2-\sigma$ constraints span $\sim 3$ orders of magnitude (green contours and
cross-hatched regions), though still rule-out large regions of currently
permitted parameter space.

In the event that the lowest frequency feature, turning point B, is not
detected, we instead arrive at the constraints shown in Figure
\ref{fig:BCDinfo_hifreq}. Provided that turning points C and D are both still
detected, we obtain the blue contours, which are broader by $\sim 1$ order of
magnitude in each dimension except for $\xiUV$, though they still close within
the broad space defined by our priors. If only the emission maximum is
detected, we instead derive the contours in green. $\Tmin$ is unconstrained in
this scenario, and only limits are available for each parameter when
marginalizing over the others. 

The results shown in Figures \ref{fig:BCDinfo_lowfreq} and
\ref{fig:BCDinfo_hifreq} are almost certainly optimistic since it is the
spectral structure which makes signal extraction possible in the first place.
With a narrow-band measurement of the global 21-cm signal that only includes
two features, we should then expect the errors on the positions of those
turning points to be \textit{larger} than those quoted in Table
\ref{tab:reference_model}. Even so, such constraints would still be a big step
forward, ruling out large regions of currently permitted parameter space and
providing priors for next-generation experiments.

%%
% Further intepretation
%%
\subsection{Assumptions Underlying the Physical Model} \label{sec:assumptions}
Our constraints on the four-parameter model of course assume that this model
is ``correct,'' i.e., its parameters are assumed to be physically meaningful.
In this section, we describe the assumptions and approximations at the heart
of this model and the circumstances in which they may deteriorate. This will
provide a basis for our final piece of discussion regarding the use of
independent constraints and model selection techniques, to follow in
\S\ref{sec:selection}.

%%
% MODEL COMPLEXITY
%%
\subsubsection{The Star Formation History}
Our $\fcoll$-based recipe for the global 21-cm signal is certainly not unique
in its ability to model the first galaxies and the high-$z$ IGM. For example,
it would not be unreasonable to employ a more flexible ``multi-population''
model \citep[e.g.,][]{Furlanetto2006} in which the $\Lya$, soft UV, and X-ray
backgrounds are produced by distinct sources, whose redshift evolution, photon
production rates, and/or spectral energy distributions are allowed to be
different. This approach may be warranted, given that the radiative properties
and formation efficiencies (with time) of PopII and PopIII stars are expected
to be different.

Some authors have instead used empirical constraints on the SFRD at
high-$z$ to model the global 21-cm signal \citep{Yajima2015}. While in
principle such models are capable of more varied star-formation histories than
our own, and can more seamlessly be compared to pre-existing empirical
constraints on the SFRD in the post-EoR Universe (from which such SFRD models
were first born), they have more free parameters and potentially obfuscate the
dominant mode of star formation, which is of primary interest in this study.
It would be straightforward to generalize our code to test
empirically-calibrated parameterizations, which have the greatest strength at
the lowest redshifts ($z \lesssim 10$), thus complementing the $\fcoll$
approach, which is likely most accurate at the highest redshifts.

Such changes to the underlying model would prevent some of the analysis so far
presented. For example, our constraints on the stellar IMF and escape fraction
relied on the assumption of a single population well described by
time-independent values of $\fstar$, the IMF (which we model implicitly
through $\Nlw$ and $\Nion$), and $\fesc$. Such analyses could still be applied
for a single-population model with an empirical SFRD, but for any kind of
multi-population model, Equation \ref{eq:stellar_pop} no longer applies. In
addition, the value of $\Nion/\Nlw$ may take on a new meaning, since it could
probe $\Nlw$ of PopIII stars that induce turning point B, and the $\Nion$ of
more ordinary PopII star-forming galaxies responsible for driving turning
point D. 

%%
% SPS models
%%
\subsubsection{Stellar Population Models}
Even with perfect knowledge of the SFRD, properly interpreting $\Nion/\Nlw$ in
terms of the stellar population requires robust predictions from synthesis
codes, which aim to generate model SEDs as a function of time. Despite a
long history and plenty of observational datasets to compare to, such codes
are still being revised to account for updates in e.g., stellar atmosphere
models \citep[e.g., line blanketing and NLTE effects ][]{Pauldrach2001} and
evolutionary tracks \citep{Ekstrom2012,Georgy2013}. Such changes are pertinent
to the reionization community, given their effects on the ionizing photon
production efficiency per unit star formation
\citep{Leitherer2014,Topping2015}. However, further revision of stellar population synthesis models may not change how we model the global 21-cm signal, but rather change how we interpret constraints on the parameters of our model, particularly $\xiLW$ and $\xiUV$. We will revisit this point in \S\ref{sec:selection}.

%%
% Further intepretation
%%
\subsubsection{Stellar Remnants and X-ray Emission}
A complete stellar synthesis code would model the remnants of stars, in
addition to stars themselves, if a comparison to datasets in the X-ray band
were of interest. This is because neutron stars and black holes, when in
binary systems, are known to dominate the X-ray luminosity of star-forming
galaxies \citep[without active nuclei;][]{Grimm2003,Gilfanov2004,Mineo2012a},
while supernovae can provide yet another source of X-rays, either via inverse
Compton scattering off hot electrons in the remnant \citep{Oh2001}, or
indirectly by mechanically heating the interstellar medium (ISM) which then
emits thermal bremsstrahlung radiation \citep{Mineo2012b}. While we are not in
the business of comparing model and measured X-ray spectra, we are concerned
with modeling the X-ray SED of galaxies insofar as it affects the thermal
history of the IGM.

The modeling of compact object populations has become a growing industry in
recent years, motivated in large part by the development of gravitational wave
observatories, continued interest in ultra-luminous X-ray sources
\citep{Belczynski2002,Belczynski2008}, and the likely importance of compact
objects in reheating of the high-$z$ IGM
\citep{Power2009,Mirabel2011,Power2013,Fragos2013}. As in the case of
stellar population modeling, the number of compact objects and their mass
distribution is expected to depend strongly on the metallicity. Unfortunately,
the observational data is sparse, especially at low metallicities, making it
difficult to calibrate the models to local analogs of high-$z$ galaxies.

Whereas our forecast for the stellar IMF and escape fraction relied on the
assumption of time-independent (but free to vary) values for $\fstar$ and
$\Tmin$, our ability to constrain $f_X$ was intimately linked to the precise
value of $\fstar$. Without more robust predictions for the X-ray yields of
stellar populations, interpretation of $f_X$ will hinge on assumptions, or
hopefully independent constraints, on the efficiency of star formation in
high-$z$ galaxies. Even if $L_X$-SFR does not evolve much with redshift,
uncertainties in the SED of X-ray sources will cloud inferences drawn from the
global 21-cm \citep{Mirocha2014}. Disentangling the normalization and spectral
shape of the X-ray background will likely require independent measurements of
the 21-cm power spectrum \citep{Pritchard2007,Pacucci2014}.

%%
% Cosmology & the HMF
%%
\subsubsection{Cosmology and the Mass Function of Dark Matter Halos}
We have fixed cosmological parameters as well as parameters governing the halo
mass function, adopting the most up-to-date values from \textit{Planck} and
the \citet{Sheth1999} form of the mass function throughout \citep[computed
using \textsc{hmf-calc};][]{Murray2013_HMFCALC}. Variations in the
cosmological parameters alone should be a secondary effect to all
astrophysical processes we consider, but potentially discernible with
observations of the dark ages ($\nu \sim 20$ MHz), prior to first-light.
Variations in the cosmological parameters will also influence the abundance of
halos, though discrepancies in the halo mass function in the literature are
known to be primarily due to differences in calibration of the fitting
functions rather than uncertainties in cosmological parameters
\citep{Murray2013}, at least at low redshifts ($z \lesssim 2$). Calibration of
the mass function at high redshifts and for low-mass halos in which the first
objects form is limited given the dynamic range needed to resolve small halos
in large volumes. If the mass function at $z \gtrsim 20$ deviates
significantly from the \citet{Sheth1999} form, it would certainly affect the
way we interpret $\Tmin$, and thus should be considered an important avenue
for future work.

%%
% MODEL ITSELF
%%
\subsubsection{The Two-Zone IGM Formalism}
Our entire procedure hinges on the ability to rapidly generate model
realizations of the global 21-cm signal, which has led us to a simple
two-phase IGM formalism rather than more detailed (and expensive) numerical or
semi-numerical simulations. Whereas simple models have been compared to
numerical simulations in the context of the 21-cm power spectrum
\citep{Zahn2011}, and found to agree quite well, no such comparison has been
conducted for global models. As far as we can tell, this is because there has
yet to be a single numerical simulation capable of self-consistently
generating a synthetic global 21-cm signal. Doing so will require high dynamic
range, capable of resolving the first star-forming halos, the radiation
backgrounds they seed, in a domain large enough to be considered a global
volume element.

Without a suite of numerical simulations to calibrate against, we have not
attempted to attach any intrinsic uncertainty associated with our model, as
was done recently by \citet{Greig2015} in the context of the 21-cm power
spectrum. However, we do expect this formalism to be accurate over nearly the
entire redshift range covered by our calculations (i.e., we do not use it
solely out of computational necessity). The two-zone formalism operates best
when HII regions are distinct and have sharp edges, and the heating and $\Lya$
is well-modeled by a uniform background. At turning point D, overlap between
bubbles is likely minimal given that the volume filling factor of HII regions
is small ($Q \sim 0.2$). In addition, their edges are likely sharp since $f_X$
is at most $\sim$ a few, in which case X-ray binaries do not enhance
ionization and heating by much on small scales \citep{Knevitt2014}. As a
result, we do not have reason to suspect a breakdown in the formalism, at
least for the reference model we have chosen.

%%
% MODEL SELECTION
%%
\subsection{Priors and the Prospects for Model Selection} \label{sec:selection}
Changes to the physical model, like those discussed in the previous section,
generally fall into two categories: they either (1) change how we interpret
the constraints on model parameters of interest, or (2) fundamentally change
the characteristics of the modeled signal. For example, improvements to
synthesis models of stars and black holes will change how the $\xi$ parameters
relate to the stellar IMF and properties of stellar remnants, and thus change
how we interpret $\xi$ values. But, so long as we still employ the
four-parameter model, our constraints on the values of $\xi$ will \textit{not}
change. If instead we introduced new parameters that allowed $\xi$ or $\Tmin$
to evolve with redshift, we have then enhanced the flexibility of the model
enough that we may now be capable of generating realizations of the global
21-cm that our previous approach simply could not have.

A ``double reionization'' scenario, which could lead to two emission features
rather than our single ``turning point D,'' is an implausible
\citep{Furlanetto2005} but illustrative example to consider in this context.
Our four-parameter model simply could not produce two emission features. One
could imagine less drastic changes that might still have new and potentially
discernible effects on the signal through modulations of its shape, such as
redshift-dependent $\xi$ and $\Tmin$, feedback, and/or multiple distinct
source populations.

We should expect that more complex model parameterizations will only have an
easier time fitting the turning points, and thus a fit to the turning points
alone may not enable one to constrain additional parameters. Use of a more
complex parameterization may still be warranted if independent constraints on
one or more of the model parameters are available to be used as priors in the
fit. However, if we do a ``single-stage'' fit, in which we fit a physical
model directly to the data rather than using a computationally inexpensive
intermediary to extract the turning points, we may find that a more complex
model is required by the data. In order to justify the additional parameters
rigorously, more advanced inference tools are required \citep[e.g.,
\textsc{multinest},\textsc{PolyChord};][]{Feroz2009,Handley2015} to compute
the Bayesian evidence.

To date, there has only been one paper on model selection for global 21-cm
datasets \citep{Harker2015}. The evidence is expensive enough to compute that
\citet{Harker2015} was limited to relatively low-dimensional spaces and
simplistic signal models. In the future, such tests will be required in order
to test whether or not more complex models (i.e., those more complex than our
four-parameter reference model) are required by the data. This presents a
unique and challenging problem for ongoing and upcoming experiments and their
associated signal extraction pipelines.

%%
% CONCLUSIONS
%%
\section{Conclusions}
This work represents the first attempt to forecast constraints on
astrophysical parameters of interest from mock observations of the global
21-cm signal. There is clearly much still to be learned, even from synthetic
datasets, about how observations in $(\nu, \dTb)$ space translate to
constraints on the properties of the IGM and/or the properties of high-$z$
galaxies. Assuming an idealized instrument, signal recovery consistent with
the values quoted in Table \ref{tab:reference_model}, and the validity of our
four-parameter model for the global 21-cm signal, we find that:
\begin{enumerate}
    \item Constraints on the turning points constrain the model parameters well (to $\sim 0.1$ dex each, on average), with factor of $\sim 2$ improvements within reach of experiments viewing multiple sky regions and employing more complex foreground parameterizations (Figure \ref{fig:triangle_1_100_4par}). Such measurements would simultaneously constrain the ionization and thermal state of the IGM (Figures \ref{fig:igm_C}-\ref{fig:igm_D}), perhaps most notably providing strong evidence for the beginning of the EoR at $z \sim 12$.
    \item Our fiducial realization of the signal is inconsistent with star-formation in halos with virial temperatures below $\sim 10^{3.5}$ K at the 2-$\sigma$ level for the most pessimistic signal extraction scenario we consider. Such constraints are enabled in large part by a broad-band measurement of the signal, since $\Tmin$ affects all three turning points in the $\sim 40-120$ MHz interval (Figures \ref{fig:param_study}, \ref{fig:fcoll_2panel}, and \ref{fig:triangle_1_100_4par}).
    \item In the simplest case, in which all model parameters are assumed to be constant in time, we can provide limits on both the escape fraction and the stellar IMF, primarily ruling out scenarios in which UV photons originate in extreme environments with very top-heavy IMFs or very high escape fractions (Figure \ref{fig:stellar_pop}). 
    \item Our constraints on X-ray sources are comparable to those achieved at $z \lesssim 4$ via stacking and the cosmic X-ray background, though at $z \sim 20$ to which the aforementioned techniques are insensitive (Figure \ref{fig:bh_pop}). In the absence of independent information, this constraint requires an assumption about the star formation efficiency and X-ray SED of galaxies, however.
    \item With only a subset of the turning points, constraints on our reference model are considerably worse (Figures \ref{fig:BCDinfo_lowfreq} and \ref{fig:BCDinfo_hifreq}). The lowest frequency features (turning points B and C) hold the most power to constrain $\Tmin$, which will make it difficult to constrain $\Tmin$ and $\xiUV$ with observations confined to the highest frequencies. Isolated detection of the absorption feature is the most valuable single-point measurement, as it leads to confidence contours which close over the prior space, except in the case of $\xiUV$.
\end{enumerate}

The authors would like to thank Abhi Datta and Brian Crosby for useful
conversations, and the anonymous referee for many insightful comments that helped improve this paper. J.M. acknowledges support through the NASA Earth and Space
Science Fellowship program, grant number NNX14AN79H. G.J.A.H is supported
through the People Program (Marie Curie Actions) of the European Union's
Seventh Framework Program (FP7/2007-2013) under REA grant agreement no.
327999. The authors also wish to acknowledge funding through the LUNAR
consortium, which was funded by the NASA Lunar Science Institute (via
Cooperative Agreement NNA09DB30A) to investigate concepts for astrophysical
observatories on the Moon, and additional support provided by the Director’s
Office at the NASA Ames Research Center (grant number NNX15AD20A). This work
utilized the Janus supercomputer, which is supported by the National Science
Foundation (award number CNS-0821794) and the University of Colorado Boulder.
The Janus supercomputer is a joint effort of the University of Colorado
Boulder, the University of Colorado Denver and the National Center for
Atmospheric Research.

\bibliography{references}

\end{document}

%% file: macros.tex
\newcommand*{\dt}[1]{%
  \accentset{\mbox{\large\bfseries .}}{#1}}
\newcommand*{\ddt}[1]{%
  \accentset{\mbox{\large\bfseries .\hspace{-0.25ex}.}}{#1}}

\newcommand{\BLAH}{{\color{red} BLAH}}

% Journals
%\newcommand{\apj}{{\it ApJ}}
%\newcommand{\aj}{{\it AJ}}
%\newcommand{\apjs}{{\it ApJS}}
%\newcommand{\apjl}{{\it ApJL}}
%\newcommand{\mnras}{{\it MNRAS}}
\newcommand{\mnrasl}{{\it MNRASL}}

\newcommand{\paper}{{\it PAPER}}
\newcommand{\hst}{{\it HST}}
\newcommand{\jwst}{{\it JWST}}
\newcommand{\edges}{{\it EDGES}}
\newcommand{\dare}{{\it DARE}}
\newcommand{\scihi}{{\it SCI-HI}}
\newcommand{\emcee}{\textsc{emcee}}

% Physical constants
\newcommand{\kB}{k_{\text{B}}}

% Cosmology
\newcommand{\Omnow}{\Omega_{\text{m},0}}
\newcommand{\Obnow}{\Omega_{\text{b},0}}
\newcommand{\OLnow}{\Omega_{\Lambda,0}}
\newcommand{\Om}{\Omega_{\text{m}}}
\newcommand{\Ob}{\Omega_{\text{b}}}
\newcommand{\OL}{\Omega_{\Lambda}}
\newcommand{\Hnow}{H_0}
\newcommand{\Hofz}{H(z)}

% Various temperatures
\newcommand{\Tcmb}{T_{\gamma}}
\newcommand{\Tcmbnow}{T_{{\gamma},0}}
\newcommand{\Tast}{T_{\ast}}
\newcommand{\zdec}{z_{\text{dec}}}
\newcommand{\TKdec}{T_{\text{K},\text{dec}}}
\newcommand{\TS}{T_{\text{S}}}
\newcommand{\TK}{T_{\text{K}}}
\newcommand{\Tstar}{T_{\star}}

% Hydrogen and helium ions
\newcommand{\HI}{\text{H} {\textsc{i}}}
\newcommand{\HII}{\text{H} {\textsc{ii}}}
\newcommand{\HeI}{\text{He} {\textsc{i}}}
\newcommand{\HeII}{\text{He} {\textsc{ii}}}
\newcommand{\HeIII}{\text{He} {\textsc{iii}}}

% Shortcuts for common subscripts
\newcommand{\tot}{\text{tot}}
\newcommand{\internal}{\text{int}}
\newcommand{\ioniz}{\text{ion}}
\newcommand{\rec}{\text{rec}}
\newcommand{\recA}{\text{rec,A}}
\newcommand{\recB}{\text{rec,B}}
\newcommand{\igm}{\text{igm}}
\newcommand{\heat}{\text{heat}}

% Lyman-alpha stuff
\newcommand{\Lya}{\text{Ly-}\alpha}
\newcommand{\Lyn}{\text{Ly-}n}
\newcommand{\Ly}{\text{Ly-}}
\newcommand{\LyC}{\text{LyC}}
\newcommand{\nmax}{n_{\text{max}}}
\newcommand{\frec}{f_{\text{rec}}}
\newcommand{\frecn}{f_{\text{rec}}^{(n)}}
\newcommand{\frecbar}{\overline{f}_{\text{rec}}}
\newcommand{\nuLya}{\nu_{\alpha}}
\newcommand{\nuLyb}{\nu_{\beta}}
\newcommand{\nuLL}{\nu_{\text{LL}}}

% Densities
\newcommand{\nH}{n_{\text{H}}}
\newcommand{\nHe}{n_{\text{He}}}
\newcommand{\nbar}{\bar{n}^0}
\newcommand{\nHbar}{\bar{n}_{\text{H}}^0}
\newcommand{\nHebar}{\bar{n}_{\text{He}}^0}
\newcommand{\nbbar}{\bar{n}_{\text{b}}^0}
\newcommand{\rhobbar}{\bar{\rho}_{\text{b}}^0}

% Column densities
\newcommand{\NH}{N_{\text{H}}}
\newcommand{\NHe}{N_{\text{He}}}
\newcommand{\NHI}{N_{\text{H } \textsc{i}}}
\newcommand{\NHeI}{N_{\text{He } \textsc{i}}}

% Emissivities and such
\newcommand{\Jalpha}{J_{\alpha}}
\newcommand{\Jhatnu}{\widehat{J}_{\nu}}
\newcommand{\eheat}{\upepsilon_{\heat}}
\newcommand{\ecomp}{\upepsilon_{\text{comp}}}
\newcommand{\eint}{e_{\internal}}
\newcommand{\eX}{\upepsilon_X}
\newcommand{\eion}{\hat{\upepsilon}_{\text{ion}}}
\newcommand{\ealpha}{\hat{\upepsilon}_{\alpha}}
\newcommand{\enu}{\hat{\upepsilon}_{\nu}}
\newcommand{\epshat}{\widehat{\upepsilon}}
\newcommand{\enuprime}{\hat{\upepsilon}_{\nu^{\prime}}}
\newcommand{\fabs}{f_{\text{abs}}}

\newcommand{\alphaMG}{\alpha_{\text{MG}}}

\newcommand{\taubar}{\overline{\tau}_{\nu}}

% Species fractions
\newcommand{\xHI}{x_{\text{H } \textsc{i}}}
\newcommand{\xHII}{x_{\text{H } \textsc{ii}}}

% Species fractions
\newcommand{\xHeI}{x_{\text{He } \textsc{i}}}
\newcommand{\xHeII}{x_{\text{He } \textsc{ii}}}
\newcommand{\xHeIII}{x_{\text{He } \textsc{iii}}}

\newcommand{\xibar}{\overline{x}_i}

% Ionization & Recombination coefficients
\newcommand{\ionHI}{\Gamma_{\text{H } \textsc{i}}}
\newcommand{\ionHeI}{\Gamma_{\text{He } \textsc{i}}}
\newcommand{\ionHeII}{\Gamma_{\text{He } \textsc{ii}}}
\newcommand{\ionsecHI}{\gamma_{\text{H } \textsc{i}}}
\newcommand{\ionsecHeI}{\gamma_{\text{He } \textsc{i}}}
\newcommand{\ionsecHeII}{\gamma_{\text{He } \textsc{ii}}}
\newcommand{\ioncollHI}{\beta_{\text{H } \textsc{i}}}
\newcommand{\ioncollHeI}{\beta_{\text{He } \textsc{i}}}
\newcommand{\ioncollHeII}{\beta_{\text{He } \textsc{ii}}}
\newcommand{\recHII}{\alpha_{\text{H } \textsc{ii}}}
\newcommand{\recHeII}{\alpha_{\text{He } \textsc{ii}}}
\newcommand{\recHeIII}{\alpha_{\text{He } \textsc{iii}}}

% Heating rate coefficients
\newcommand{\heatHI}{\mathcal{H}_{\text{H } \textsc{i}}}
\newcommand{\heatHeI}{\mathcal{H}_{\text{He } \textsc{i}}}
\newcommand{\heatHeII}{\mathcal{H}_{\text{He } \textsc{ii}}}

% Cooling rate coefficients
\newcommand{\cooldielHeII}{\omega_{\text{He } \textsc{ii}}}

% Number densities of common ions
\newcommand{\nHI}{n_{\text{H } \textsc{i}}}
\newcommand{\nHII}{n_{\text{H } \textsc{ii}}}
\newcommand{\nHeI}{n_{\text{He } \textsc{i}}}
\newcommand{\nHeII}{n_{\text{He } \textsc{ii}}}
\newcommand{\nHeIII}{n_{\text{He } \textsc{iii}}}
\newcommand{\nel}{n_{\text{e}}}  
\newcommand{\ntot}{n_{\text{tot}}}

% 21-cm features
\newcommand{\zB}{z_{\text{B}}}
\newcommand{\zC}{z_{\text{C}}}
\newcommand{\zD}{z_{\text{D}}}
\newcommand{\nuB}{\nu_{\text{B}}}
\newcommand{\nuC}{\nu_{\text{C}}}
\newcommand{\nuD}{\nu_{\text{D}}}
\newcommand{\TB}{\delta T_b (\nu_{\text{B}})}
\newcommand{\TC}{\delta T_b (\nu_{\text{C}})}
\newcommand{\TD}{\delta T_b (\nu_{\text{D}})}
\newcommand{\znull}{z_{\text{null}}}
\newcommand{\ztrans}{z_{\text{trans}}}
\newcommand{\zfl}{z_{\ast}}
\newcommand{\zbh}{z_{\bullet}}
\newcommand{\zrei}{z_{\text{rei}}}

% BH stuff
\newcommand{\acc}{\text{acc}}
\newcommand{\fduty}{f_{\text{duty}}}
\newcommand{\Cedd}{C_{\text{edd}}}
\newcommand{\tedd}{t_{\text{edd}}}
\newcommand{\fedd}{f_{\text{edd}}}
\newcommand{\Mbh}{M_{\bullet}}
\newcommand{\Mdot}{\dot{M}}
\newcommand{\MdotBH}{\dot{M}_{\bullet}}
\newcommand{\mdot}{\dot{m}}
\newcommand{\Ledd}{L_{\text{edd}}}
\newcommand{\BHeff}{\zeta_{\text{acc}}}
\newcommand{\fX}{f_X}
\newcommand{\fXeff}{f_{X,\mathrm{eff}}}
\newcommand{\fsc}{f_{\text{sc}}}

% Random
\newcommand{\xtot}{x_{\tot}}
\newcommand{\coll}{\text{coll}}
\newcommand{\zprime}{z^{\prime}}
\newcommand{\fstar}{f_{\ast}}
\newcommand{\fstarbh}{\tilde{\fstar}}
\newcommand{\fbh}{f_{\bullet}}
\newcommand{\fcoll}{f_{\text{coll}}}
\newcommand{\fcollprime}{f_{\text{coll}}^{\prime}}
\newcommand{\dfcolldz}{\frac{df_{\text{coll}}}{dz}}
\newcommand{\dfcolldztwo}{\frac{d^2f_{\text{coll}}}{dz^2}}
\newcommand{\dfcolldt}{\frac{df_{\text{coll}}}{dt}}
\newcommand{\dfcolldzprime}{\frac{df_{\text{coll}}^{\prime}}{dz}}
\newcommand{\dfcolldtprime}{\frac{df_{\text{coll}}^{\prime}}{dt}}
\newcommand{\dfcolldzbh}{\frac{d\tilde{f}_{\text{coll}}}{dz}}
\newcommand{\dfcolldtbh}{\frac{d\tilde{f}_{\text{coll}}}{dt}}
\newcommand{\mmin}{m_{\text{min}}}
\newcommand{\rhobh}{\rho_{\bullet}}
\newcommand{\rhobhdot}{\dt{\rho}_{\bullet}}
\newcommand{\rhobhdotacc}{\dt{\rho}_{\bullet, \mathrm{acc}}}
\newcommand{\rhobhdotnew}{\dt{\rho}_{\bullet, \mathrm{new}}}
\newcommand{\rhobhdotejec}{\dt{\rho}_{\bullet, \mathrm{ejec}}}
\newcommand{\rhobhdottot}{\dt{\rho}_{\bullet, \tot}}

\newcommand{\rhostar}{\rho_{\ast}}
\newcommand{\rhostardot}{\dt{\rho}_{\ast}}
\newcommand{\rhostarbhdot}{\dt{\rho}_{\ast\bullet}}
\newcommand{\rhom}{\rho_m}
\newcommand{\rhobbarnow}{\bar{\rho}_b^0}
\newcommand{\rhombarnow}{\bar{\rho}_m^0}
\newcommand{\fstardegen}{f_{\ast \bullet}}
\newcommand{\Nion}{N_{\text{ion}}}
\newcommand{\Nlw}{N_{\text{LW}}}
\newcommand{\Nalpha}{N_{\alpha}}
\newcommand{\fesc}{f_{\text{esc}}}
\newcommand{\Msun}{M_{\odot}}
\newcommand{\Tvir}{T_{\text{vir}}}
\newcommand{\Tmin}{T_{\text{min}}}
\newcommand{\Tminprime}{T_{\text{min}}^{\prime}}

\newcommand{\nHbarnow}{\bar{n}_{\text{H}}^0}

\newcommand{\fion}{f_{\text{ion}}}
\newcommand{\nnu}{$n_{\nu}$}
\newcommand{\ncol}{N_i}

\newcommand{\zpeak}{z_{\text{peak}}}

\newcommand{\bol}{\mathrm{bol}}
\newcommand{\fbol}{f_{\bol}}
\newcommand{\nuHub}{\nu_{\mathrm{Hub}} }

\newcommand{\fheat}{f_{\text{heat}}}
\newcommand{\fXh}{f_{X,h}}
\newcommand{\fioni}{f_i^{\text{ion}}}
\newcommand{\Lbol}{\mathcal{L}_{\text{bol}}}
\newcommand{\spec}{\mathcal{N}}
\newcommand{\Heat}{\mathcal{H}}
\newcommand{\trec}{$t_{\text{rec}}$}
\newcommand{\Lbox}{L_{\mathrm{box}}}
\newcommand{\dx}{\Delta x}
\newcommand{\dd}{\text{d}}

\newcommand{\drIF}{$\Delta r_{\mathrm{IF}}$}
\newcommand{\dTb}{\delta T_b}
\newcommand{\dTbdot}{\dot{\delta T_b}}
\newcommand{\Nvec}{\mathbf{N}}
\newcommand{\sh}{\mathrm{sh}}

% Units
\newcommand{\sfrd}{\Msun \ \mathrm{yr}^{-1} \ \mathrm{cMpc}^{-3}}
\newcommand{\intensityunitsnumber}{\text{s}^{-1} \ \text{cm}^{-2} \ \mathrm{Hz}^{-1} \ \text{sr}^{-1}}
\newcommand{\intensityunitsenergy}{\text{erg} \ \text{s}^{-1} \ \text{cm}^{-2} \ \mathrm{Hz}^{-1} \ \text{sr}^{-1}}
\newcommand{\coemissivityunitsnumber}{\text{s}^{-1} \ \mathrm{Hz}^{-1} \ \text{cMpc}^{-3}}
\newcommand{\coemissivityunitsenergy}{\text{erg} \ \text{s}^{-1} \ \mathrm{Hz}^{-1} \ \text{cMpc}^{-3}}

\newcommand{\emissivityunitsnumber}{\text{s}^{-1} \ \mathrm{Hz}^{-1} \ \text{cMpc}^{-3}}
\newcommand{\emissivityunitsenergy}{\text{erg} \ \text{s}^{-1} \ \mathrm{Hz}^{-1} \ \text{cMpc}^{-3}}

\newcommand{\coheatingdensity}{\text{erg} \ \text{s}^{-1} \ \text{cMpc}^{-3}}
\newcommand{\coenergydensity}{\text{erg} \ \text{cMpc}^{-3}}

\newcommand{\cXunits}{\text{erg} \ \text{s}^{-1} \ (\Msun \ \text{yr})^{-1}}

\newcommand{\dprime}{\prime\prime}
\newcommand{\zdprime}{z^{\dprime}}

\newcommand{\Nsky}{N_{\text{sky}}}
\newcommand{\tint}{t_{\text{int}}}
\newcommand{\xiLW}{\xi_{\text{LW}}}
\newcommand{\xiXR}{\xi_X}
\newcommand{\xiUV}{\xi_{\text{ion}}}
\newcommand{\QHII}{Q_{\textsc{HII}}}